\documentclass[aps,prb,preprint,nofootinbib,preprintnumbers,eqsecnum,floatfix,longbibliography,11pt,noshowpacs]{revtex4-2}
\usepackage{amsmath,amssymb,physics,mathtools}
\usepackage[dvipsnames]{xcolor}
\usepackage{graphicx}
\usepackage{siunitx}
\usepackage{physics}
\AtBeginDocument{\RenewCommandCopy\qty\SI}

\usepackage{printlen}
\usepackage{hyperref}
\hypersetup{colorlinks=true,linkcolor=blue,citecolor=red,urlcolor=blue}

\usepackage[nohyperlinks,nolist]{acronym}

\definecolor{forest}{rgb}{0.13, 0.55, 0.13}

\newcommand{\thetaHall}{\theta_{\text{Hall}}}
\newcommand{\tauHall}{\tau_{\text{Hall}}}
\newcommand{\taumomrel}{\tau_{\text{mom.rel.}}}
\newcommand{\tauimp}{\tau_{\text{imp}}}
\newcommand{\bmu}{\bar{\mu}}

\newcommand{\non}{\nonumber\\}

\newcommand{\eps}{\epsilon}

\newcommand{\bx}{\mathbf{x}}
\newcommand{\by}{\mathbf{y}}
\newcommand{\bk}{\mathbf{k}}

\newcommand{\cnn}{\chi_{nn}}
\newcommand{\cns}{\chi_{ns}}
\newcommand{\css}{\chi_{ss}}
\newcommand{\cen}{\chi_{\epsilon n}}
\newcommand{\cee}{\chi_{\epsilon \epsilon}}

\newcommand{\ve}{\varepsilon}
\newcommand{\dTF}[1]{\delta T_{(#1)}}
\newcommand{\dmuF}[1]{\delta \mu_{(#1)}}
\newcommand{\dvF}[1]{\delta v_{(#1)}}
\newcommand{\dvB}[1]{\delta \bar{v}_{(#1)}}

\newcommand{\muh}[1]{\hat \mu^{(#1)}}
\newcommand{\chipp}{{\chi_{\pi\pi}}}

\newcommand{\omegaAfour}{{\omega_{A,(4)}}}
\newcommand{\barv}[2]{\overline{#1}_{(#2)}}

\newcommand{\susone}[2]{{\frac{\partial #1}{\partial #2}}}

\newcommand{\sustwom}[2]{{\frac{\partial^2 #1}{\partial #2^2}}}
\newcommand{\susthreem}[2]{{\frac{\partial^3 #1}{\partial #2^3}}}

\newcommand{\dta}[2]{\Delta^{#1}_{#2}}

\newcommand{\omegaccan}{{\omega^{\text{can}}_c}}
\newcommand{\omegacobs}{{\omega^{\text{obs}}_c}}
\newcommand{\omegapcan}{{\omega_{p,\text{can}}^2}}

\DeclareMathOperator{\realPart}{Re}
\DeclareMathOperator{\imagPart}{Im}

\usepackage{tabularx}
\begin{document}

\begin{acronym}
  \acro{rn}[RN]{Reissner-Nordstr\"{o}m}
  \acro{gr}[GR]{Gubser-Rocha}
\end{acronym}

\count\footins = 1000
\author{N. Chagnet}
\author{S. Arend}
\author{F. Balm}
\author{M. Janse}
\author{J. Saldi}
\author{K. Schalm}
\affiliation{Instituut-Lorentz for Theoretical Physics, $\Delta$-ITP, Leiden
University, The Netherlands.}
\email{kschalm@lorentz.leidenuniv.nl}

\title{
  Natural
  anomalous
  cyclotron
  response in a
  hydrodynamic
local quantum critical  metal in a periodic potential.}

\begin{abstract}
  We study DC magnetotransport in a quantum critical metal in the presence of a
  lattice. In the regime where
  the transport is hydrodynamical the interplay of the
  Lorentz force and the lattice gives rise to a natural
  anomalous contribution
  to the cyclotron frequency that changes it from its canonical charge-to-mass
  ratio.
  The size of this effect is universal as it is determined only by
  thermodynamic quantities.
  Remarkably the Drude weight changes in such a way that to first subleading
  order in the lattice strength the Hall resistivity and Hall coefficient do not
  change, though the Hall angle does change.
  We confirm our results with numerical simulations in a holographic model of a
  strange metal. For weak lattice strength these hydrodynamic effects are shown
  to be present.
  The numerical simulations also suggest that strong lattice effects beyond a
  hydrodynamic regime may provide a
  resolution to the experimentally observed
  anomalous Hall response of cuprate strange metals.
  \\[.2in]
  {\em Jan Zaanen passed away during the research stage of this article. We
  dedicate it to his memory.}
\end{abstract}
\maketitle
\newpage
\section{Introduction}
\label{sec:introduction}

The nature of ``strange metals'' as realized in the strongly interacting
electron systems of condensed matter physics has been one of the most pressing
questions since the late 1980s. Soon after the discovery of superconductivity
at a high temperature in copper oxides, it was found that the metallic state
above the superconducting transition is characterized by highly anomalous
properties that are very different from those of regular Fermi-liquid metals.
The accumulation of experimental information on cuprates and related systems
since then increasingly fortified the notion that a completely different
physical principle is at work. Transport properties play here an important
role. The linear-in-temperature DC electrical resistivity down to the lowest
temperatures is often taken as the defining characteristic of a strange metal.
It suggests a momentum relaxation time $\rho_{xx} =\frac{1}{\sigma_{xx}} \sim
\frac{1}{\tau_{\text{mom.rel.}}}$  that is set by a Planckian dissipation
scale $\tau_{\hbar} = \hbar / (k_B T)$ \cite{zaanenWhyTemperatureHigh2004}.
On general grounds it can be argued that this represents a fundamental bound
under linear response conditions for the rate of thermalization that can only
be reached in a state characterized by dense many-body entanglement
\cite{zaanenPlanckianDissipationMinimal2019,hartnollPlanckianDissipationMetals2022}.

The other famous experimental anomaly in the context of cuprate strange metals
is the temperature dependence of the Hall angle. This refers to the
observation that the ratio of the longitudinal linear conductivity to the Hall
conductivity scales quadratically in the temperature
$\frac{\sigma_{xx}}{\sigma_{xy}}=\frac{\rho_{xx}}{\rho_{yx}} \equiv
\cot(\thetaHall) \sim T^2$
\cite{chienUnusualMathitTemperature1991,riceHallEffectMathit1991,chienEffectZnImpurities1991}.
In regular Fermi-liquid metals the Hall angle is controlled by the same
(linear-momentum-relaxation) timescale $\cot(\thetaHall) \sim
\frac{1}{\tau_L}=\frac{1}{\tau_{\text{mom.rel.}}}$. Even if the temperature
scaling of the momentum-relaxation timescale is anomalous, one would therefore
expect the same anomalous temperature scaling in the Hall angle. This is not
seen in strange metals, however. This suggests a second ``Hall relaxation
timescale'' $\tau_H \sim \frac{1}{T^2}$ associated with the Lorentz force
exerted by an external magnetic field that behaves in a different manner than
the one governing the resistivity $\tau_{L} \sim 1/T$
\cite{andersonHallEffectTwodimensional1991}, originating perhaps in different
even-odd charge-conjugation scattering rates (\emph{e.g.},
\cite{colemanHowShouldWe1996}) or spin charge separation of the electron in a
two fluid spinon-holon model (\emph{e.g.},
\cite{leeTransportPhenomenologyHolonspinon1997}).

The necessity of such a second timescale effectively rules out that a strange
metal can be described by a Fermi liquid or a variant thereof. For small
magnetic fields $B \sim \SI{1}{\tesla}$, the cyclotron frequency of a free
quasiparticle $\omega_c = q B/m \sim 10^{11}~\unit{\per\second}$ is much
smaller than the disorder/lattice scattering rate $\Gamma_L\equiv \tau_L^{-1}
\sim v_F/a_{\text{Lattice}} \sim %
10^{15}~\unit{\per\second}$.\footnote{In practice the cyclotron frequency is
  used as a precise way to measure the (averaged band-) mass of the
  quasiparticle. In Fermi Liquids which flows to a free fixed point in the IR,
  this is the only relevant operator and all other quantities are then a
  function of this cyclotron mass $m_c$. Within perturbation theory this mass is
  between 1\textendash{}2 times the free electron mass $m_e =
\SI{0.511}{\mega\eV}$ and does not affect the above argument.}
The dissipation underlying the Hall angle response of an electron travelling in
a Lorentz orbit can then be viewed as successive piecewise linear resistance
and will therefore be governed by the momentum relaxation scattering
rate.\footnote{This same insight is at the origin of Kohler's rule in the
  longitudinal magnetoresistance
\mbox{$(\rho_{xx}(B)-\rho_{xx}(0))/\rho_{xx}(0)=f(\omega_c/\rho_{xx}(0))$}.}

The simplicity of this argument illustrates that the introduction of a second
timescale is a tall order for {\em any} quasiparticle-like model in an atomic
lattice.\footnote{A possibility is
  \cite{varmaPRL2001,varmaPRL2001Erratum,abrahamsHallEffectMarginal2003}. The
  subtlety of the specific effect that it relies on illustrates the main point
above.}
Put more concretely: any Boltzmann-transport theory of charged quasiparticles
where the slowest single relaxation time (from electron-electron interactions)
is parametrically larger than the lattice size, will have a Hall resistance
set by a piecewise linear resistance (from Umklapp/disorder scattering) for
magnetic field strengths smaller than the inverse lattice size $B< mv_F/q
a_{\text{L}} \sim \SI{e4}{\tesla}$.
Fermi-surface scattering with anisotropic rates will not help this
\cite{colemanHowShouldWe1996}.\footnote{
  A recent study in the longitudinal magnetoresistance $\rho_{xx}(B)$ does
  attempt such an anisotropic quasiparticle scattering model for LSCO and
  Nd-LSCO with reasonable fit to the data up to $T\sim \SI{30}{\kelvin}$
  \cite{grissonnancheLinearinTemperatureResistivity2021a}, but it does not fit
  well with other high $T_c$ superconductors
  \cite{berbenCompartmentalizingCuprateStrange2022}. Reasons why anisotropic
  quasiparticle scattering is unlikely to settle the issue are (1) that the
  anisotropy scale of the Fermi-surface is never larger than the lattice size.
  Therefore momentum relaxation due to disorder is always the parametrically
  larger scale, unless the sample is ultra-clean. However, doped cuprates always
  have significant disorder. (2) Only at low temperatures does Fermi surface
  anisotropy have a significant effect in scattering (see \emph{e.g.},
  \cite{cookElectronHydrodynamicsPolygonal2019,bakerNonlocalElectrodynamicsUltrapure2023a}).
  At higher temperatures thermal broadening washes out the anisotropy. And
  (3) any anisotropic $T$-linear scattering will dominate piece-wise linear over
  any $T^2$-component at low $T$. Disorder scattering can subsequently
  ``isotropize'' this $T$-linear channel. Anisotropic Fermi liquid quasiparticle
  scattering is therefore unlikely to explain the observed cuprate Hall angle
scaling seen up to $T\sim \SI{200}{\kelvin}$.}
The most likely remedy, especially if the explanation is to be semi-universal
to explain the same phenomenon in multiple different cuprates, is that the
sparsely entangled quasiparticle picture must be abandoned. There are other
experimental signatures, notably the plasmon-width
\cite{nuckerPlasmonsInterbandTransitions1989,mitranoAnomalousDensityFluctuations2018},
the single-fermion spectral-width
\cite{varmaPhenomenologyNormalState1989,reberUnifiedFormLowenergy2019,Smit:2021dwh},
the non-quasiparticle-SYK-explanation of the $\omega^{-2/3}$ scaling in the
optical conductivity
\cite{patelUniversalTheoryStrange2022,michonPlanckianBehaviorCuprate2022a,liStrangeMetalSuperconductor2024},
and recent shot noise measurements
\cite{chenShotNoiseIndicates2022,nikolaenkoTheoryShotNoise2023}, that also
point to the conclusion that cuprate strange metal transport is not
explainable through weakly interacting quasiparticles.

The discovery of the holographic AdS/CFT correspondence and subsequent
rediscovery of the Sachdev-Ye-Kitaev quantum spin liquid to describe densely
entangled many body theories with non-quasiparticle transport
\cite{zaanenHolographicDualityCondensed2015,hartnollHolographicQuantumMatter2018,zaanenLecturesQuantumSupreme2021a,sachdevQuantumStatisticalMechanics2023}
allows us to address this question of Hall transport in the absence of
quasiparticles theoretically. In both approaches an underlying quantum
critical state protects the dense entanglement and the absence of
quasiparticle-excitations. At the same time these critical fixed points are
strongly interacting and cannot be described with conventional perturbative
quantum critical approaches. We shall take the holographic approach to study
magnetotransport; an equivalent result is expected from an SYK computation as
its ground state is equivalent to the universal AdS$_2$ ground state found in
holographic models at finite chemical potential
\cite{kitaevKITPlecture2015,maldacenaCommentsSachdevYeKitaevModel2016,maldacenaConformalSymmetryIts2016}
(see
  \cite{sachdevUniversalLowTemperature2019,sachdevQuantumStatisticalMechanics2023}
for a review); the SYK approach to magnetotransport is studied in
\cite{guoCyclotronResonanceQuantum2023} with qualitatively similar results to
what we describe below.\footnote{The non-hydrodynamic regime of strong
  disorder is studied from the SYK perspective in
\cite{patelMagnetotransportModelDisordered2018}.}  Two important points must
be emphasized at the outset. (1) To have any finite transport translational
symmetry must be broken: the lattice and/or disorder must be added by hand in
both SYK and holographic models. (2) In the absence of quasiparticles the
natural language of transport is not the Boltzmann equation, but
hydrodynamics. Importantly, correlated with the absence of quasiparticles, the
collectivization scale where hydrodynamics sets in is extremely small both in
SYK and in holographic models. The lattice or disorder scale is generically
larger than this collectivization scale. Many of the SYK-AdS$_2$ results can
therefore be phenomenologically understood in terms of hydrodynamics in the
presence of translational symmetry breaking
\cite{davisonHolographicDualityResistivity2014,lucasConductivityStrangeMetal2015,lucasHydrodynamicTransportStrongly2015,hartnollTheoryUniversalIncoherent2015,donosDiffusionInhomogeneousMedia2017},
though there is a distinction between strong and weak momentum relaxation; see
\cite{zaanenHolographicDualityCondensed2015,hartnollHolographicQuantumMatter2018}
for reviews.\footnote{Within the context of holographic duality for strongly
  correlated systems that are two specific scenarios for thermo-electric
  transport that have been explored in detail. One are so-called massive
  gravity/axion/Q-lattice models: these are based on a single momentum
  relaxation rate without sacrificing homogeneity \emph{e.g.},
  \cite{andradeSimpleHolographicModel2013,donosHolographicQlattices2014,kimCoherentIncoherentMetal2014,blakeMomentumRelaxationFluid2015}.
  These do not capture spatially dependent scattering/gradient contributions
  such as here, or effects from Umklapp modes as in
  \cite{balmTlinearResistivityOptical2023}, and cannot explain magnetotransport in the cuprates \cite{ahnHolographicGubserRochaModel2023}. The other is translational symmetry
  breaking through charge density waves; see \emph{e.g.},
  \cite{delacretazTheoryHydrodynamicTransport2017,baggioliColloquiumHydrodynamicsHolography2023a}.
  These break translations spontaneously whose additional Goldstone current has
its own particular physics that can differ from plain hydrodynamics \cite{amorettiHydrodynamicMagnetotransportCharge2021,amorettiHydrodynamicMagnetotransportHolographic2021,armasApproximateSymmetriesPseudoGoldstones2022,armasHydrodynamicsPlasticDeformations2023a}.}

Hydrodynamic magneto-transport of a densely entangled non-quasiparticle theory
in the presence of weak disorder was already studied in one of the first
applications of holography to strongly correlated electron systems
\cite{hartnollTheoryNernstEffect2007}. For relativistic hydrodynamics these
authors indeed found a second relaxation timescale:
$\gamma = \sigma_Q B^2/(\eps+P)$. This parity-even timescale predominantly
affects the longitudinal magneto-resistance and originates in the
intrinsic diffusive contribution to the charge current that does not contribute
to momentum flow %
with strength
parametrized by the phenomenological transport coefficient $\sigma_Q$. This
transport coefficient equals $\sigma_Q = \frac{T}{\mu^2}\bar{\kappa}_Q$ where
$\bar{\kappa}_Q$ is the anomalous heat conductivity which in a regular Fermi
liquid originates from the Lindhard continuum (\emph{e.g.}, Appendix D of
\cite{lucasElectronicHydrodynamicsBreakdown2018}). In the non-relativistic
limit $\mu=m_ec^2 \rightarrow \infty$ and this contribution from $\gamma$
becomes negligible.

In relativistic quantum critical systems a non-vanishing $\sigma_Q$ which
contributes predominantly to $\rho_{xx}\sim \frac{1}{\sigma_Q} $ opens up the
theoretical possibility that at low $T$ this momentum-relaxation independent
part sets the observed linear-in-$T$ scaling, and the momentum-relaxation rate
sets the Hall resistivity $\rho \sim \frac{1}{\taumomrel}$
\cite{blakeQuantumCriticalTransport2015}. In such a scenario one can obtain a
reasonable hydrodynamical explanation of DC magnetotransport in the cuprates
\cite{amorettiHydrodynamicalDescriptionMagnetotransport2020}. By not including
the optical response, however, this analysis misses the fact that for most of
the observed temperature regime the longitudinal optical conductivity
$\sigma(\omega)$ shows a clear Drude peak (\emph{e.g.},
\cite{vanheumenStrangeMetalElectrodynamics2022,michonPlanckianBehaviorCuprate2022a}).
In a finite density (\emph{i.e.}, finite doping) system it is hard to think of
a scenario where this Drude peak and hence the DC value of resistivity is not
controlled by momentum relaxation, in which case the Hall conductivity must
originate in a second time-scale. A thorough understanding of cuprate
transport must therefore include the optical finite frequency response.

Our first main result is
the demonstration
that magneto-transport of a densely entangled quantum critical
non-quasiparticle theory in the presence of a weak lattice
has new additional non-dissipative contributions to transport that are more
important than the relaxation timescale $\gamma$.
Such qualitatively terms were first identified in \cite{armasApproximateSymmetriesPseudoGoldstones2022,armasHydrodynamicsPlasticDeformations2023a} in the context of charge density waves and their importance in magnetotransport
was recently noted in
\cite{gouterauxDrudeTransportHydrodynamic2023}, which appeared while this work
was being completed.
The most obvious consequence of these additional contributions is that the interplay of the perpendicular magnetic field and
the 2D lattice results in an anomalous shift in the cyclotron frequency:
$\omegacobs = \omegaccan + A^2 \omega_A$ with $A$ the strength of the lattice
potential.\footnote{The appearance of this shift $\omega_A$ can be
  misinterpreted as a second parity-odd non-dissipative timescale, as we shall
  discuss, but it is not.
}
Its main origin can
be qualitatively understood
as arising from the variance contribution to cyclotron frequency when
considered as
ratio of static charge-momentum and momentum-momentum-susceptibilities
$\omega_c = \frac{\chi_{\pi j}}{\chi_{\pi\pi}}B$. In a translationally invariant
relativistic system these equal $\chi_{\pi j}=n$ and $\chi_{\pi\pi}=\eps+P$, or
in non-relativistic system $\chi_{\pi j}=n$ and
$\chi_{\pi\pi}=nm$,\footnote{Note that
  we use natural units where the unit of electric charge $q=1$ and speed of
  light
$c=1$.} but formally susceptibilities are two point functions
$\omega_c = \frac{\langle j \pi\rangle}{\langle \pi\pi\rangle} B$.
In a weak periodic translational symmetry breaking potential, where
thermodynamics applies locally $n(\bx) =\bar{n}+ \Delta n(\bx)$, there are
hydrodynamic cross correlations $\langle \Delta n(\bx)\Delta\pi(\by)\rangle
\neq 0$ that can now contribute to the spatial average (zero momentum) of the
susceptibility
(Section \ref{sec:magneto-hydro}).
The exact expressions
for $\omega_c$ and $\omega_A$
are more involved and their derivation from dissipative magneto-hydrodynamics
is summarized in Appendix~\ref{appendix:magneto-hydro}.
It is natural to presuppose that this shift is the hydrodynamic counterpart of
the cyclotron frequency corrections due to Umklapp or another form of
translational symmetry breaking
\cite{kankiTheoryCyclotronResonance1997,guoCyclotronResonanceQuantum2023} and
might therefore be absorbed in a renormalization of $\chi_{\pi\pi}=\eps+P$
which is the relativistic hydrodynamic generalization of  the effective (band)
mass $m$ (times the density),
but this is not the case, as we shall show.
What it does signal, is that spatial averages that are measured in experiment
can no longer be readily combined for different observables.

At the same time there is a remnant of the robustness of magneto-transport to
distortions in line with Kohn's and Kohler's theorems. But instead of the
cyclotron frequency, this is evident in the Hall resistivity.
Even in the presence of a weak lattice the Hall resistivity remains equal to
ratio of the shifted cyclotron frequency and the plasmon frequency squared
(the weight of the Drude peak) $\rho_{yx} = \frac{\omega_c}{\omega_p^2}$. The
plasmon frequency also shifts, however, and in such a way that remarkably to
first subleading order
the simple expression $\rho_{yx} = B/\bar{n}$ still holds, despite the
dissipative and non-dissipative shift in both the cyclotron  and the plasmon
frequency.
Though we will not confirm this quantitatively, at least qualitatively this
opens up a theoretical possibility of a cyclotron mass that evolves with
doping while the Hall resistivity stays conventional as observed in cuprate
strange metals \cite{legrosEvolutionCyclotronMass2022}. The observational consequences of these two results are summarized in Section \ref{sec:relevancy}.

In Section \ref{sec:bh-rn-numerics} we confirm this cyclotron frequency shift
but nevertheless unchanged Hall resistivity due to hydrodynamic charge
transport in the presence of a lattice
by observing this phenomenology
in
numerical computations of magneto-transport in the holographic
\ac{rn} AdS$_2$ model for a strange metal in the presence of a periodically modulated
chemical potential representing the atomic lattice.
At long time scales and long distances hydrodynamics emerges directly from this
computational model and is not input. For weak lattices the results match
seamlessly with our analytical predictions; they also match direct numerical
hydrodynamical simulations in the presence of a lattice that include Umklapp
\cite{chagnetHydrodynamicsRelativisticCharged2023}.
At the same time the \acl{rn} AdS$_2$ model does not have the
correct scaling  properties to be  a good candidate to explain the
phenomenology of the cuprates, even though it is a densely entangled strange
metal state. The premier candidate with the appropriate scaling properties is
the so-called \ac{gr} model which has the correct $T$-linear resistivity
and a $T$-linear specific heat. The power of the universal phenomenological
magneto-hydrodynamical description is that we can immediately predict the Hall
response in the \acl{gr} model in the presence of a weak lattice. We
present it in Section \ref{sec:gubser-rocha} not only for completeness, but
also to note the possibility of a curious cancellation where a formally
subleading timescale can become the dominant one.

By their very nature, perturbative weak lattice effects are still controlled by
a large single relaxation time dominating over smaller secondary relaxation
times. The weak lattice scenario can never explain the observed cuprate Hall
anomaly in the cuprates. Our second main result is observational: the
numerical \acl{rn} simulations also allow us to probe the strong
lattice regime. As the lattice strength increases one clearly sees the
validity of single time-scale physics fail. The Hall resistivity notably has a
qualitatively different temperature dependence compared to the longitudinal
resistivity. The unique insight that the holographic numerical simulations
give, allow us to disentangle its origin. In strong lattice effects the Hall
coefficient $R_H=\rho_{yx}/B$ is no longer inversely proportional to the
average charge density. It directly implies that measurements of the Hall
coefficient in the cuprates when interpreted as effective density should be
handled with care.
In Section \ref{sec:strong-lattice-and-cuprate-hall} we discuss the possible
relevancy of our results with respect to experiment and
we conclude with a consideration whether large lattice strengths do have the
potential to explain the similarly observed physics in the cuprates.

\section{Magneto-hydrodynamic-transport in a weak periodic potential: a
cyclotron frequency shift and other timescales}
\label{sec:magneto-hydro}

We first discuss magneto-hydrodynamic transport in the weak lattice regime both
to exhibit the novel hydrodynamic response, but also  to validate our later
numerical simulations and provide confidence that extrapolation to large
lattices is reliable.
The numerically studied holographic \acl{rn} system shows emergent
relativistic hydrodynamics; we therefore restrict our discussion to that here.
For completeness the full derivation in Appendix \ref{appendix:magneto-hydro}
also gives the results for non-relativistic hydrodynamics.

The hydrodynamic description of transport in a weak periodic potential/weak
disorder is a theoretically coherent extension of the Drude model under the
condition that the lattice periodicity/disorder length is larger than the mean
free path that determines the onset of hydrodynamics and local thermodynamic
equilibrium
\cite{lucasHydrodynamicTransportStrongly2015,lucasTransportInhomogeneousQuantum2016,lucasHydrodynamicsElectronsGraphene2018}.
Then, given the dominant channel by which the translational symmetry breaking
is communicated, hydrodynamics provides a consistent way to compute the matrix
of relaxation times
$\tau_{ij}$ that relates the emergent spatially averaged velocity of the
charged fluid as response to a static %
electric field:
\begin{align}
  \label{eq:tau-def}
  \overline{\chipp} ~ (\tau^{-1})_{ij} \bar{v}^j =
  (\bar{n}\delta_{ik}+\sigma_QB_{ik})E^k
\end{align}
Here $\overline{\chipp} = \bar \eps + \bar P$ is the momentum susceptibility
and $\bar{\eps}, \bar{P}$ are the spatially averaged background energy and
pressure respectively --- as indicated above we use relativistic hydrodynamics.
In a weakly broken background all these quantities are spatially dependent with
a perturbative expansion $\eps(\bx)=\bar{\eps}+A\hat{\eps}(\bx) + \ldots $
controlled by the lattice amplitude or disorder strength $A$. An important
aspect will be that in perturbation theory the spatial average itself can
contain higher order (even power) corrections in $A$:
$\bar{\eps}=\barv{\eps}{0} + \underbracket[0.140ex]{\barv{\eps}{2}}_{\sim A^2}
+\ldots$, even though experiments will only measure the sum of all these
contributions in the total \mbox{average $\bar{\eps}$} \textemdash{} we will
refer to these as hydrostatic corrections. Within hydrodynamics the
fundamental relation of hydrodynamics is obeyed {\em locally} $\eps(\bx)+ P
(\bx)=s(\bx)T+\mu(\bx) n(\bx)$.
We shall consider two-dimensional systems only, \emph{i.e.}, with indices
$i,j,k,\ldots \in \{1,2\}$, but use notation where the (spatially constant)
magnetic field perpendicular to the two-dimensional system $B_{ij} =
\eps_{ijz}B^z=B\eps_{ij}$ is an in-plane two tensor. $\sigma_Q$ is the
previously mentioned microscopic transport coefficient that appears in the
constitutive relation
for the charge current
\begin{align}
  \hspace{-.2in}J_i(\bx,t) & = n(\bx,t) v_i(\bx,t) -
  \sigma_Q\!\left(\!T(\bx,t)\partial_i\frac{\mu(\bx,t)}{T(\bx,t)}-
  E_i(\bx,t)-B_{ij}(\bx,t)v^j(\bx,t)\!\right)
\end{align}
such that it allows for current flows with no net momentum flow. Such a term is
notably important in systems that are near charge neutrality, such as graphene
\cite{fritzQuantumCriticalTransport2008a,muellerQuantumcriticalRelativisticMagnetotransport2008,schuettCoulombInteractionGraphene2011,lucasHydrodynamicsElectronsGraphene2018}.\footnote{Note
  that in a relativistic system with Lorentz symmetry, the other microscopic
  thermoelectric coefficients relating charge and heat transport are all
  constrained by the value of $\sigma_Q$. The more general non-relativistic
  derivation in Appendix \ref{appendix:magneto-hydro} includes the coupling to
  the heat transport as allows for arbitrary microscopic coefficients $\alpha_Q,
\bar{\kappa}_Q$ in thermo-power and heat transport.}

The conductivity tensor $\sigma_{ij}$, defined through
$\bar{J}_{i}=\sigma_{ij}E^j$, follows from using Eq.~\eqref{eq:tau-def} in the
spacetime independent part \textemdash{} the spatial average\footnote{Note
  that for periodic perturbations, the spatial average will be assumed to take
  the form $\int \dd^2 \bx = \Bigl( \frac{G}{2\pi} \Bigr)^2
\int_{-\pi/G}^{\pi/G} \dd x \dd y$.} \textemdash{} of the linearized current
fluctuation \cite{lucasHydrodynamicTransportStrongly2015}
\begin{align}
  \label{eq:constitutive-current}
  \hspace{-.2in}    \bar{J}_i & = \bar{n} \bar{v}_i +
  \sigma_Q(E_i+B_{ij}\bar{v}^j) +{\sigma_Q}\!\!\int\!\!\text{d}^2\bx
  \frac{\mu(\bx)}{T}\partial_i\delta T(\bx;\bar{v},E,B) +
  \!\!\int\!\!\text{d}^2\bx (n(\bx)-\bar{n}) v_i(\bx;\bar{v},E,B) .
\end{align}
The final two terms arise from the fact that in linear response the velocity-
and temperature-fluctuations $v_i(\bx,\bar{v}), \delta T(\bx,\bar{v})$ can be
spatially varying, even if the background temperature and steady state
velocity do not.
Ignoring these two terms, \emph{i.e.}, assuming the chemical potential and
density to be spatially constant, the charge current is then simply $\bar J_i
= \bar n \bar v_i + \sigma_Q (E_i + B_{ij} \bar v^j)$ and, assuming that to
leading order there is no interplay between the magnetic field and momentum
relaxation,
an analysis of the momentum current gives the inverse relaxation time matrix
$\tau^{-1}$ as \cite{hartnollTheoryNernstEffect2007}
\begin{align}
  \tau^{-1} =
  \begin{pmatrix}
    \Gamma_{\text{imp}}+\gamma & -\omega^{\text{can}}_c     \\
    \omega^{\text{can}}_c      & \Gamma_{\text{imp}}+\gamma
  \end{pmatrix}
\end{align}
with $\Gamma_{\text{imp}}=\frac{1}{\tauimp}$ a Drude impurity relaxation rate,
$\gamma=\sigma_Q B^2/(\bar{\eps}+\bar{P})$ the timescale of
\cite{hartnollTheoryNernstEffect2007} mentioned in the introduction, and the
canonical cyclotron frequency $\omega^{\text{can}}_c=
\bar{n} B/(\bar{\eps}+\bar{P})$. Then solving Eq.~\eqref{eq:tau-def} for
$\bar{v}^j$ and substituting on the right hand side of the constitutive
relation Eq.~\eqref{eq:constitutive-current}, one obtains an Ohms's law
$\bar{J}_i = \sigma_{ij}E^j$ with the DC conductivity tensor of
\cite{hartnollTheoryNernstEffect2007}
\begin{align}
  \label{eq:hartn-sach-cond}
  \hspace*{-.3in}  \sigma_{ij} & =\frac{1}{\bigl( \Gamma_{\text{imp}} + \gamma
  \bigr)^2 + (\omegaccan)^2} \!
  \begin{pmatrix}
    \Gamma_{\text{imp}} + \gamma & \omegaccan                   \\
    -\omegaccan                  & \Gamma_{\text{imp}} + \gamma
  \end{pmatrix}\!\! %
  \begin{pmatrix}
    \omega_{p,\text{can}}^2 - \sigma_Q \gamma     & 2 \sigma_Q \frac{B \bar
    n}{\overline \chipp} \\
    -2 \sigma_Q \frac{B \bar n}{\overline \chipp} & \omega_{p,\text{can}}^2 -
    \sigma_Q \gamma
  \end{pmatrix} + \sigma_Q \delta_{ij} \nonumber\\ %
  & =
  \sigma_Q\frac{1}{(\Gamma_{\text{imp}}+\gamma)^2+(\omega^{\text{can}}_c)^2}\!
  \begin{pmatrix}
    \Gamma_{\text{imp}}(\Gamma_{\text{imp}}+\gamma+\frac{(\omegaccan)^2}{\gamma})
    &
    \omega^{\text{can}}_c(2\Gamma_{\text{imp}}+\gamma+\frac{(\omega^{\text{can}}_c)^2}{\gamma})
    \\
    -\omega^{\text{can}}_c({2}\Gamma_{\text{imp}}+\gamma+\frac{(\omega^{\text{can}}_c)^2}{\gamma})
    &
    \Gamma_{\text{imp}}(\Gamma_{\text{imp}}+\gamma+\frac{(\omegaccan)^2}{\gamma})
  \end{pmatrix}.
\end{align}
where the (canonical) plasma frequency (or Drude weight) equals $\omegapcan =
{\bar n}^2/\overline \chipp$ and in the last step repeated use is made of the
identity $\omegapcan=\frac{\sigma_Q(\omega^{\text{can}}_c)^2}{\gamma}$.
In the weak momentum relaxation limit $\Gamma_{\text{imp}}=1/\tauimp
\rightarrow 0$ this is equivalent to the resistivities\footnote{The
  theoretical  computation below, as well as numerical computations, naturally
  yield conductivities through the Kubo relation. To compare directly with
  experiment which measures resistivities (voltage response to supplied current,
  see \emph{e.g.}, \cite{ayresIncoherentTransportStrange2021}), and to avoid the
  confusion of an apparent insulating regime in $\sigma_{xx} \sim
  \frac{1/\tau_0}{\frac{1}{\tau_0^2}+\omega_c^2}$ when $\omega_c\gg {1/\tau_0}$,
we have chosen to present the equivalent resistivities.}
\begin{align}
  \label{eq:07-resistivities}
  & \rho_{xx}
  = \frac{\Gamma_{\text{imp}}}{\omega_p^2}
  -\frac{\gamma\Gamma_{\text{imp}}^2}{\omega_p^2
  (\gamma^2+(\omegaccan)^2)}+\ldots\non
  & \rho_{yx}
  = \frac{\omegaccan}{\omega_p^2}-\frac{\sigma_Q
  \omegaccan}{\omega_p^2}\frac{\gamma\Gamma_{\text{imp}}^2}{\omega_p^2(\gamma^2+(\omegaccan)^2)}+\ldots
\end{align}
which reduce to the standard expression in the non-relativistic limit
$\mu=m_ec^2\rightarrow \infty$ where the microscopic transport coefficient
$\sigma_Q=\frac{T}{\mu^2}\bar{\kappa}_Q$ becomes negligible, \emph{i.e.},
$\sigma_Q\rightarrow 0$ and hence $\gamma\rightarrow 0$.

The first result we report here is that a careful hydrodynamic derivation of
magnetotransport where translational symmetry breaking is imprinted through a
locally varying external chemical potential
$\mu_{\text{ext}}(\bx)=\bmu+A\hat{\mu}_{\text{ext}}(\bx)$, and therefore
includes all the terms in Eq.~\eqref{eq:constitutive-current}, gives, firstly,
an expression for $\tau^{-1}$ of the following form\footnote{The result for
  the longitudinal magnetoresistance was first derived by this method in
  \cite{baumgartnerMagnetoresistanceRelativisticHydrodynamics2017}.
}
instead:
\begin{align}
  \label{eq:tau-exact}
  \tau^{-1} =
  \begin{pmatrix}
    \Gamma_{(2)}+\Gamma_{(4)}+\gamma~~  & -\omega^{\text{can}}_c- \omegaAfour \\
    \omega^{\text{can}}_c+\omegaAfour~~ & \Gamma_{(2)}+\Gamma_{(4)} +\gamma
  \end{pmatrix}~ + {\cal O} (A^6,A^4B,A^2B^2,B^3)
\end{align}
In the setting with weak translational symmetry breaking ($A \ll 1$)  at a
scale larger than the onset of hydrodynamics ($\ell_\text{mom.rel.} \gg
\ell_\text{m.f.p.}$) and weak magnetic fields ($B \sim A^2$) the new terms
$\Gamma_{(4)}\sim A^4$ and $\omegaAfour \sim A^2B$ in the expression above are
of equal order with respect to the correction due to $\gamma \sim B^2$. For
small magnetic fields the leading term is the cyclotron frequency shift
$\omegaAfour$. This shift will show up directly in the dynamical response: as
we derive in Appendix \ref{appendix:magneto-hydro}
the true cyclotron frequency as measured in an optical conductivity experiment
---such as
\cite{postObservationCyclotronResonance2021,legrosEvolutionCyclotronMass2022}
--- is $\omegacobs=\omegaccan+ \omegaAfour$, with $\omegaAfour$ the anomalous
shift with respect to the canonical value.\footnote{Unlike in
  \cite{hartnollTheoryNernstEffect2007} one cannot deduce this from the
  extension $\tau^{-1}\rightarrow \tau^{-1}-i\omega\delta_{ij}$ in
  Eq.~\eqref{eq:tau-def} and applying the subsequent steps. See Appendix
\ref{appendix:magneto-hydro}.}

Secondly, with the proper inclusion of the second term in
Eq.~\eqref{eq:constitutive-current}
the longitudinal and Hall resistivities are
\begin{align}
  \label{eq:full-resist}
  \rho_{xx} & = \frac{1}{\omegapcan}\left(\Gamma_{(2)}  +
  \Gamma_{(4)}-\frac{\sigma_Q\Gamma_{(2)}^2}{\omegapcan} + \rho_{(4)}\right)
  + {\cal O} (A^6,A^2B^2) \non
  \rho_{yx} & =  \frac{B}{\bar{n}} + {\cal O} (A^4B,B^3)
\end{align}
where $\rho_{(4)}$ is an extra contribution at order $A^4$ which we will explain shortly.
The fact that the expression for the Hall resistivity $\rho_{yx}$ remains
unchanged to first order despite the shift in the cyclotron frequency relies
on a remarkable identity
\begin{align}
  \int\!d^2\bx (n(\bx)-\bar{n}) v_i(\bx,\bar{v})+ \int\!d^2\bx
  \frac{\mu(\bx)}{T}\partial_i\delta T(\bx,\bar{v}) =  \bar{n}
  \frac{\omegaAfour}{\omega^{\text{can}}_c}+\ldots
\end{align}
There is a priori no apparent reason why the corrections to the cyclotron
frequency $\omega$ do not contribute to the Hall resistivity at that order,
although
the recent results of \cite{gouterauxDrudeTransportHydrodynamic2023}, indicated
a possible mechanism why this might happen. We shall explain further below.

\bigskip

The full computation of $\omegacobs, \tau^{-1}$ and the conductivities
$\rho_{yx}$ and $\rho_{xx}$ is involved, and reported in Appendix
\ref{appendix:magneto-hydro}. A crucial aspect is that a perturbative approach
to weak lattice magneto-transport that has no order-of-limits problem only
holds if the magnetic field $B \sim \eps^2$ scales quadratically with the
lattice strength $A\sim \eps$ for the small parameter $\eps$ (see Eq.~(21) in
\cite{Lucas:2016omy}).
A second crucial aspect is that one must consider the full AC response rather
than limiting to time-independent quantities from the beginning. The final
result is of the same form as Eq.~\eqref{eq:tau-exact} and
Eq.~\eqref{eq:hartn-sach-cond}
\begin{align}
  \label{eq:true-cond}
  \sigma_{ij} & =\frac{1}{\bigl( \Gamma + \gamma \bigr)^2 + (\omegacobs)^2}
  \begin{pmatrix}
    \Gamma + \gamma & \omegacobs      \\
    -\omegacobs     & \Gamma + \gamma
  \end{pmatrix} \cdot
  \begin{pmatrix}
    \omega_{p}^2 - \sigma_Q \gamma           & 2 \sigma_Q \frac{B n}{\overline
    \chipp} \\
    -2 \sigma_Q \frac{B n}{\overline \chipp} & \omega_{p}^2 - \sigma_Q \gamma
  \end{pmatrix} + \sigma_Q \delta_{ij}
\end{align}
but with
\begin{align}
  \label{eq:final-cyclotron-drude}
  \Gamma   = \Gamma_{(2)}+\Gamma_{(4)}~,\quad \gamma = \gamma_{(4)}=
  \frac{\barv{\sigma_Q}{0} B^2}{\barv{\chipp}{0}}~, \quad \omegacobs =
  \frac{\barv{n}{0} +  \barv{n}{2} + 2 \lambda_{n,(2)}}{\barv{\chipp}{0} +
  \barv{\chipp}{2} + \lambda_{\pi,(2)}}  B~,
\end{align}
But this is not the only change. Importantly, the Drude weight changes as well
\begin{align}
  \label{eq:drude-weight-correction}
  \omega_p^2 = \frac{\Bigl(\barv{n}{0}  + \barv{n}{2}  +
  \lambda_{n,(2)}\Bigr)^2}{\barv{\chipp}{0} +  \barv{\chipp}{2}  +
  \lambda_{\pi,(2)}}
\end{align}
This corrected Drude weight $\omega_p^2 = \omegapcan + \omega_{p,(2)}^2$ is the origin of the extra contribution to the longitudinal resistivity $\rho_{(4)}$ in Eq.~\eqref{eq:full-resist}.
The various parts are particular combinations of averaged thermodynamic
quantities as reflected in Table \ref{tab:summary}.
They are of two types: the hydrostatic corrections account for  higher order
contributions to the average density from a locally varying chemical potential
$\bar{n}=\barv{n}{0}+\barv{n}{2}+\ldots$, ${\overline \chipp}=\barv{\chipp}{0}
+  \barv{\chipp}{2}+\ldots$, and are simply higher order susceptibilities
\begin{align}
  \label{eq:higher-order-suscep}
  \barv{n}{2} = \frac{1}{2}A^2\int\!\dd^2\bx\,
  \hat{\mu}(\bx)\hat{\mu}(\bx)\frac{\partial^2}{\partial \mu^2} \barv{n}{0}  =
  \frac{1}{2}\frac{\partial}{\partial \mu} \barv{\chi_{nn}}{0}
  |\hat{\mu}_{\text{ext}}(\bk)|^2~,~\text{etc}
\end{align}
{where $|\hat{\mu}_{\text{ext}}(\bk)|^2 \equiv \sum_{\bk}
  \hat{\mu}_{\text{ext}}(-\bk) \hat{\mu}_{\text{ext}}(\bk)$.
}
The coefficients $\lambda_n$ and $\lambda_\pi$, however, denote additional
non-hydrostatic corrections
on top of the expected hydrostatic corrections.
Such corrections $\lambda_i$ were shown to naturally arise when momentum
relaxation is mediated by a massless scalar operator in
\cite{armasApproximateSymmetriesPseudoGoldstones2022,armasHydrodynamicsPlasticDeformations2023a,gouterauxDrudeTransportHydrodynamic2023}, building on charge density wave studies
\cite{armasHydrodynamicsChargeDensity2020,amorettiHydrodynamicMagnetotransportHolographic2021,amorettiHydrodynamicsDimensionalStrongly2022b}. There the extra conserved
current associated to that operator allows for new transport coefficients in
the hydrodynamics constitutive relations that map one-to-one to the terms
$\lambda_i$ above. There are no new conserved currents in our weak lattice
set-up mediated through a spatially modulated chemical potential. Nevertheless
the corrections we have derived in Appendix \ref{appendix:magneto-hydro} can
be completely recast in the same form, with $\lambda_i$ now not an independent
transport coefficient but determined by underlying thermodynamic quantities as
in Table \ref{tab:summary}.\footnote{We are very grateful to B. Gout\'eraux
  for suggesting this. See Section \ref{sec:connection-blaise-shukla} for a
detailed discussion.}
Intuitively
the gradient of the background chemical potential
indeed
plays a similar role as the scalar gradient in the scalar model, but the fact
that the formal structure of the two expressions is the same is remarkable.

\begin{table}[t!]
  \begin{align*}
    \begin{array}{|>{\scriptstyle}l|}
      \hline\\[-2em]
      {\displaystyle
        ~\Gamma_{(2)}  = {
        \frac{|\hat{\mu}_{\text{ext}}(\bk)|^2}{2\barv{\chipp}{0}^3}} \biggl[
          \frac{1}{\barv{\sigma_Q}{0}}\bigl( \barv{n}{0} \barv{\cen}{0} -
          \barv{\chipp}{0} \barv{\cnn}{0}\bigr)^2 + \barv{\cen}{0}^2
      (\barv{\eta}{0}+\barv{\zeta}{0}) G^2 \biggr]} \\[2em]
      {\displaystyle
        ~\Gamma_{(4)} ~:~ \text{See Table~\ref{tab:summary-Gamma4}, Appendix
        \ref{appendix:magneto-hydro}}.
      }\\[2em]
      \hline\\[-2em]
      {\displaystyle ~\lambda_{n,(2)}  =
        \frac{|\hat{\mu}_{\text{ext}}(\bk)|^2}{2\barv{\chipp}{0}^2} \Bigl[
          \barv{\cen}{0} \barv{\sigma_Q}{0} (\barv{\eta}{0}+\barv{\zeta}{0})
          G^2  +
          (\barv{\cen}{0} + 2 \barv{n}{0})( \barv{n}{0} \barv{\cen}{0} -
      \barv{\chipp}{0} \barv{\cnn}{0})\Bigr]~,}
      \\
      {\displaystyle~\lambda_{\pi,(2)}  =
        \frac{|\hat{\mu}_{\text{ext}}(\bk)|^2}{2\barv{\chipp}{0}} \Biggl[
          \barv{\cen}{0}^2 - 2 \barv{\cnn}{0} - \frac{\barv{\cen}{0}^2
          \barv{\cee}{0}}{\barv{\chipp}{0}^3}  (\barv{\eta}{0}+\barv{\zeta}{0})
        G^2} \\
        {\hspace{1.1in}\displaystyle - 2 \frac{(\barv{\eta}{0}+\barv{\zeta}{0})
          \barv{\cen}{0}}{\barv{\sigma_Q}{0} \barv{\chipp}{0}^3} \Bigl(
            \barv{n}{0}
          \barv{\cen}{0} - \barv{\chipp}{0} \barv{\cnn}{0}  \Bigr) \Bigl(
            \barv{n}{0}
        \barv{\cee}{0} - \barv{\chipp}{0} \barv{\cen}{0}  \Bigr)     } \\
        {\hspace{1.1in}\displaystyle       - \frac{\Bigl( \barv{n}{0}
            \barv{\cen}{0} - \barv{\chipp}{0} \barv{\cnn}{0} \Bigr)^2 \Bigl(
              \barv{n}{0}^2
              \barv{\cee}{0} - 2 \barv{n}{0} \barv{\cen}{0} \barv{\chipp}{0} +
          \barv{\cnn}{0} \barv{\chipp}{0}^2 \Bigr)}{\barv{\chipp}{0}^3
      \barv{{\sigma_Q}}{0} G^2 } \Biggr]                                       }
      \\[2em]
      \hline
    \end{array}
  \end{align*}
  \caption{\footnotesize{
      The leading correction of the momentum relaxation rate $\Gamma_{(2)}$ and
      non-hydrostatic corrections $\lambda_{i,
      (2)}$ obtained from hydrodynamics in the presence of a magnetic field and a weak
      perturbative lattice sourced by $\mu(\bx)=\bar{\mu}+ \hat{\mu}(\bx)$
      (Appendix
      \ref{appendix:magneto-hydro}).
      The expression for $\Gamma_{(4)}$ is lengthy and is  given in the
      appendix. The result presented is for a square parity-symmetric lattice
      with
      equal lattice vectors $\bk_x=G, \bk_y=G$. For the specific square lattice
      \eqref{eq:ionic-potential-cosine}, we will later use
      $|\hat{\mu}_{\text{ext}}(\bk)|^2 = \frac{\bmu^2 A^2}{4}$. Here
      $\chi_{\eps n}
      =\frac{\partial \eps}{\partial \mu}, \chi_{\eps\eps} = T \frac{\partial
      \eps}{\partial T} + \mu \frac{\partial \eps}{\partial \mu}$ are the
      energy-charge cross-susceptibility and energy-energy susceptibility
      respectively in addition to the charge-charge susceptibility
      ${\chi}_{nn}$ and
      momentum-momentum susceptibility ${\chi}_{\pi\pi}$; $\eta$ is the
      (spatially
      averaged) shear viscosity, $\zeta$ is the (spatially averaged) bulk
      viscosity,
      and $\sigma_Q$ the microscopic conductivity transport coefficient
      allowing for
      charge flow with no net momentum flow. For all quantities the overbar
      means
      spatially averaged over a unit cell, and the subscript $(n)$ indicates the
      order in $A$ contribution.  %
      \label{tab:summary}
  }}
\end{table}

\subsection{Relevancy for experiment}
\label{sec:relevancy}

From these exact results
in weak lattice/incoherent metal magnetotransport where the momentum relaxation
scale is larger than the mean-free-path,
a number of important observational consequences follow:

\begin{itemize}
  \item[1.] Observationally the %
    clearest one
    is the cyclotron frequency shift
    \begin{align}
      \label{eq:omega-c-shift}
      \omegaAfour = \frac{\barv{n}{0}}{\barv{\chipp}{0}}\biggl(
        \frac{\barv{n}{2} + 2 \lambda_{n,(2)}}{\barv{n}{0}} -
        \frac{\barv{\chipp}{2} +
      \lambda_{\pi,(2)}}{\barv{\chipp}{0}} \biggr) B.
    \end{align}
    In translationally invariant systems the cyclotron frequency does not
    change (Kohn's theorem), and this shift is therefore not unexpected even if
    the analytical form computed here was not yet known.
  \item[2.]

    Nevertheless the cyclotron frequency and the Hall resistivity are still
    related by 
    \begin{align}
        \rho_{yx} &= \frac{\omegacobs}{\omega_p^2} + \,\,{\cal O}(A^6)
    \end{align}    
    to first subleading order in the lattice strength $A$ but with the
    corrected expressions for both the cyclotron frequency {\em and} the Drude
    weight.
    A comparison between the optical response that measures $\omega_c$ and
    $\omega_p^2$ (in a Drude-like regime) and DC transport that measures
    $\rho_{yx}$ can therefore serve as an indicator as to whether weak
    lattice(/weak disorder) hydrodynamics is at work.

  \item[3.] The hydrostatic corrections to the charge density and momentum
    susceptibilities are the same in both the cyclotron frequency and the Drude
    weight but size of the non-hydrostatic corrections differ \textemdash{} note
    the extra factor of $2$ in the numerator of
    Eq.~\eqref{eq:final-cyclotron-drude}. This extra factor of $2$ is the hidden
    reason of the unexpected observation that at this order in perturbation
    theory
    the Hall coefficient $R_H=\frac{\rho_{yx}}B$ remains equal to
    $\frac{1}{\bar{n}} = \frac{1}{n_{(0)}+n_{(2)}+\ldots}$:
    \begin{align}
      R_{H} = \frac{\omegacobs}{\omega_p^2B} = \frac{\barv{n}{0} +  \barv{n}{2}
      + 2 \lambda_{n,(2)}}{\Bigl(\barv{n}{0}  + \barv{n}{2}  +
      \lambda_{n,(2)}\Bigr)^2} = \frac{1}{\bar{n}} + {\cal O}(\lambda_{n,(2)}^2)
    \end{align}
    It does suggest that at the next order such a cancellation will no longer
    happen.
    This does presuppose that in the optical response $\omega_c$ and
    $\omega_p^2$ are well defined, \emph{i.e.}, the optical response is to the
    eye
    Drude like. In bad metals/the bad metal regime, where by definition this is
    not so, this relation between the Hall coefficient and the density breaks
    down
    and should not be used: we explicitly show this break-down in numerical
    simulations below.

  \item[4.] In contrast, the longitudinal resistivity does receive
    corrections: they are a more precise version of second time scale found in
    \cite{hartnollTheoryNernstEffect2007} as shown in
    Eq.~\eqref{eq:07-resistivities}. %

    From Eq.~\eqref{eq:true-cond} one deduces
    \begin{align}
      \label{eq:shifted-hydro-relaxation-rate}
      \rho_{xx} & = \frac{\Gamma}{\omega_p^2}-
      \frac{\gamma_{(4)}(\Gamma_{(2)})^2}{\omega_p^2(\omegacobs)^2} +\ldots~. \non
      & = \frac{\Gamma}{\omega_p^2} -
      \frac{\barv{\sigma_Q}{0}(\Gamma_{(2)})^2}{\omega_p^4} +\ldots
    \end{align}
    Note that there are contributions in the first term that are of the same
    order in $A^4$
    as the second term proportional to $\gamma_{(4)}$ --- the term found in
    \cite{hartnollTheoryNernstEffect2007}. This includes in particular the
    non-hydrostatic corrections $\lambda_{n,(2)}$ and $\lambda_{\pi,(2)}$ in
    $\omega_p^2$. In the second line we have rewritten the expression to make
    clear that any reference to the role of $\omegacobs$ or dependence on the
    magnetic field $B$ is a mathematical artifact at this point: this is not a
    longitudinal magnetoresistance term, but a transport contribution from
    current
    flow without momentum.

    In the non-relativistic Galilean-invariant limit, this second term
    proportional to $\gamma_{(4)}$ vanishes and we recover $\rho_{xx} =
    \frac{\Gamma}{\omega_p^2}$ up to sub-leading order, but again with the
    corrections at first subleading order in $A^4$. %

    Whether the term proportional to $\sigma_{Q}$ and hence $\gamma_{(4)}$
    should be considered in an experimental set-up is very difficult to deduce
    of
    charge transport alone. A combination experiment which includes also the
    open
    circuit thermal conductivity $\kappa=\bar{\kappa}-T\alpha\sigma^{-1}\alpha$
    can do so in principle as has been advocated in several recent articles
    \cite{blakeDiffusionChaosAdS22017,lucasElectronicHydrodynamicsBreakdown2018,balmTlinearResistivityOptical2023}.\footnote{Using
      that the Hall coefficient continues to be equal to
      $R_H=\frac{1}{\bar{n}}$
      even in the presence of weak translational symmetry breaking, in that
      setting
      the microscopic transport coefficient $\sigma_Q$ can be extracted from
      the open boundary heat conductivity
      as $\kappa_{xx}(B=0) = \frac{1}
    {TR_H^2(\omega_p)^4\sigma_{xx}(B=0)}\sigma_Q(\sigma_{xx}(B=0)-\sigma_Q)$.}

  \item[5.] The Hall angle will have a second time-scale
    \begin{align}
      \cot(\thetaHall) = \frac{\rho_{xx}}{\rho_{yx}} = \frac{\Gamma}{\omega_c}
      - \frac{\barv{\sigma_Q}{0}(\Gamma_{(2)})^2/\omega_p^2}{\omega_c} ~
    \end{align}
    which scales as $T^2$ if the leading momentum relaxation rate scales linear
    $\Gamma_{(2)}\sim T$. However, this term is already present in the
    longitudinal resistivity $\rho_{xx}$ where it is always subleading for weak
    lattices and cannot be a hint at the explanation for the observed Hall angle
    anomaly in the cuprates. In numerical simulations of holographic strange
    metals below we shall see that for strong lattices a truly new time-scale
    will
    emerge.
  \item[6.]
    Due to the underlying hydrodynamics, all the expressions and in particular
    the corrections to Drude transport are built out of thermodynamic variables,
    susceptibilities and transport coefficients of the homogeneous
    translationally
    invariant background and these can be input from any theory/spatially
    averaged
    experiment. The result is therefore not sensitive to details of the system
    but
    only depends on the equation of state and macroscopic transport. Such
    hydrodynamic effects may provide an explanation for the universality of
    strange metal transport \cite{davisonHolographicDualityResistivity2014}. A
    similar line of reasoning was put forward without magnetic field within the
    same holographic model \cite{balmTlinearResistivityOptical2023}.
\end{itemize}
While the calculations presented here in %
relativistic hydrodynamics to be able to compare to numerical simulations in
the next sector, the %
results %
are only
marginally dependent on this assumption.
The derivation in Sec.~\ref{sec:galilean-limit} for non-relativistic
Galilean
hydrodynamics
shows that the only difference is that in almost all instances of $\sigma_Q$ in
the relaxation rate and resistivities, it simply becomes $\frac{T_0}{\bmu^2}
\bar \kappa_Q$ with $\kappa_Q$ the anomalous heat diffusion coefficient. The
notable exception is in the DC conductivity $\sigma_{xx}$ itself and the extra
diffusion term $\gamma$: the $\sigma_Q$ to the former and hence $\gamma$
itself vanish in the limit of non-relativistic hydrodynamics.

\section{Holographic strange metal models dual to AdS black holes as numerical
experiments for magnetotransport}
\label{sec:bh-rn-numerics}

\begin{figure}[t]
  \centering
  \includegraphics[width=\textwidth]{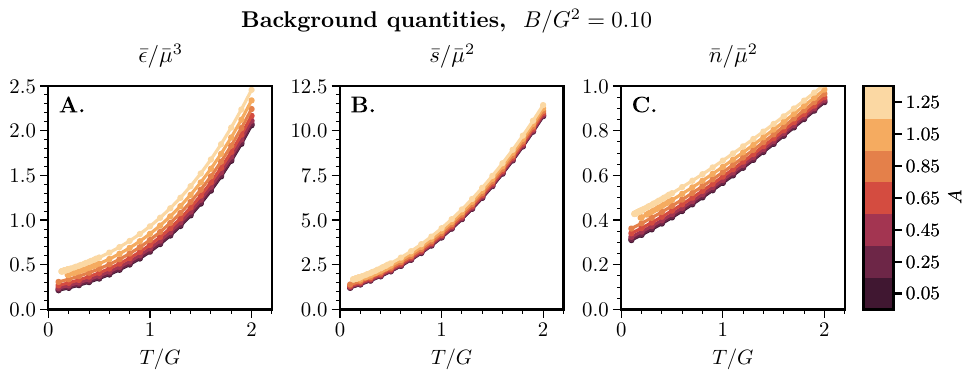}
  \caption{\footnotesize{
      The unit lattice cell spatially averaged $\bar{\eps}$, entropy density
      $\bar{s}$, charge density $\bar{n}$
      as a function of temperature in units of the lattice momentum $T/G$, for
      a 2D holographic \acl{rn} strange metal model in the presence of a
      weak
      square lattice sourced by a chemical potential
      $\mu_{\text{ext}}(x,y)=\bmu +
      \frac{\bmu A}{2}\Bigl(\cos(G x) +A \cos(G y)\Bigr)$ with strength $A$ and
      a
      perpendicular magnetic field $B$. The fixed parameters are $B/G^2 =
      \num{0.1}$
      and $G/\bmu = \num{0.1}$. All quantities are normalized by appropriate
      factors
      of $\bmu$.
  }}
  \label{fig:RN-thermodynamics}
\end{figure}

We can confirm the hydrodynamic effects we just highlighted (the
  non-hydrostatic corrections to $\omega_c$ while the Hall resistivity remains
$\rho_{yx}=\frac{\omega_c}{\omega_p^2}=\frac{B}{\bar{n}}$) by a numerical
experiment. Holographic models of strange metals describe finite density
systems without quasiparticles where the hydrodynamic regime emerges in a
single set-up
\cite{casalderrey-solana_liu_mateos_rajagopal_wiedemann_2014,zaanenHolographicDualityCondensed2015,hartnollHolographicQuantumMatter2018}.
We have computed the DC magnetotransport response of a holographic
\acl{rn} strange metal model in a perpendicular magnetic field in
the presence of an external lattice. The set-up is described in detail (with
no magnetic field) in \cite{balmTlinearResistivityOptical2023} and briefly
reviewed (with a magnetic field) in appendix \ref{app:numerics-RN}. The
external ionic lattice potential
\begin{align}
  \label{eq:ionic-potential-cosine}
  \mu(\bx)=\bmu + \frac{\bmu A}{2}\Bigl(\cos(Gx)+\cos(Gy)\Bigr)
\end{align}
is perturbative with amplitude $A \ll 1$ and lattice vector $G$ held fixed at
$\frac{G}{\bmu}=\num{0.1}$; a value $G\ll \bmu$ is necessary to ensure that
the translation symmetry breaking length scale  is larger than the onset of
hydrodynamics $\ell_{\text{mom.rel.}} \gg \ell_{\text{m.f.p.}}$ and the
formalism of hydrodynamic perturbation theory applies. In such numerical
experiments, we have the possibility to measure not only transport properties
but also the thermodynamics of the system (averaged over a unit cell); these
are presented in Fig.~\ref{fig:RN-thermodynamics}.

\begin{figure}[t]
  \includegraphics[width=\textwidth]{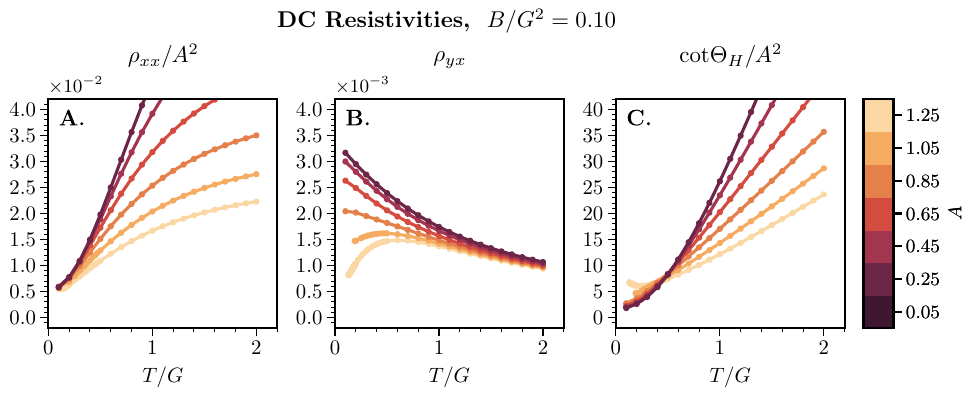}
  \caption{\footnotesize{The longitudinal and Hall DC resistivities,
      $\rho_{xx}$ and $\rho_{yx}$ and the Hall angle \mbox{$\cot \Theta_H =
      \frac{\rho_{xx}}{\rho_{yx}}$}, computed numerically for a 2D holographic
      \acl{rn} strange metal model in a periodic potential
      $\mu_{\text{ext}}(x,y)=\bmu + \frac{\bmu A}{2}\Bigl(\cos(G x) +A \cos(G
      y)\Bigr)$ with strength $A$ and perpendicular magnetic field $B$. The
      fixed
      parameters are $B/G^2 = \num{0.1}$ and $G/\bmu = \num{0.1}$.
      Note that $\rho_{xx}$ and $\cot \Theta_H$ are both normalized by the
      lattice strength.
  }}
  \label{fig:DCCondVsT}
\end{figure}

Fig.~\ref{fig:DCCondVsT} shows the results for the longitudinal DC resistivity
$\rho_{xx}=\sigma_{yy}/\det(\sigma_{ij})$ and Hall DC resistivity $\rho_{yx} =
\sigma_{xy}/\det(\sigma_{ij})$
as a function of $T/\mu$ for various lattice strengths.
Absent a magnetic field, the low energy limit of the \ac{rn} model is a quantum
critical strange metal state whose scaling properties predict a
low-temperature longitudinal DC resistivity that is constant in $T/\mu$,
$\rho_{xx}^{\text{(RN)}} \sim \frac{1}{\omega_p    ^2\tau_0} \sim T^0$
\cite{davisonHolographicDualityResistivity2014}.
The numerical data at low $A$ is over such a large temperature range that it
includes the transition to this regime: in Fig.~\ref{fig:DCCondVsT}
\mbox{$\rho_{xx} \sim \frac{1}{\omega_p^2\tau_0}\sim a_0 + a_1 T+a_2 T^2
+\ldots$}
More importantly,
Fig.~\ref{fig:DCCondVsT}.C shows how at low $A$ the Hall angle
$\cot(\thetaHall)\sim \frac{1}{\omega_c\tau_0} %
$ agrees with the $\rho_{xx}$ temperature-behavior  over essentially the full
computed $T/G$ regime, consistent with single relaxation time physics.

Crucially, we also see a deviation from this %
single relaxation time physics
as we increase $A$ where notably the cotangent of the Hall angle scales
differently and stronger in $T$ than the longitudinal resistivity $\rho_{xx}$
for the same configuration. We shall analyze this regime in Section
\ref{sec:strong-lattice-and-cuprate-hall}.
For the sake of completeness, our set-up allows for the computation of the full
longitudinal and transverse thermoelectric conductivity matrix.
The results for these are presented in Appendix \ref{app:numerics-RN}
Fig.~\ref{fig:thermoelectric-conductivities}.

\subsection{Validity of hydrodynamics for weak lattices}
\label{sec:validity-hydro}

The hydrodynamics description derived in the previous Section
\ref{sec:magneto-hydro} ought to be an accurate description of the
small-lattice strength strange metal models holographically dual to
\acl{rn} AdS black holes we have computed.
To establish
this
we need an independent determination of $\omega_p^2$, $\omega_c$ and $\sigma_Q$.
This can only be extracted reliably from an AC optical conductivity experiment.
For the RN-AdS model with a two-dimensional potential, this is at this time not
yet numerically feasible. We therefore complement our results with an AC pure
hydrodynamics computation which uses the thermodynamics and the transport
coefficients of the RN-AdS model as input. The details
are described in Appendix~\ref{appendix:magneto-hydro}. Inspired by
Ref.~\cite{lucasHydrodynamicTransportStrongly2015}, this formalism relies on a
limit in which the two cyclotron modes are the dominant modes of transport
(see the special expansion \eqref{eq:hydro-expansion-fluctuations-lucas}), but
includes the characteristic lattice Umklapp effects in hydrodynamic
perturbations explained in
\cite{balmTlinearResistivityOptical2023,chagnetHydrodynamicsRelativisticCharged2023}.

\begin{figure}[t!]
  \centering
  \includegraphics[width=0.95\textwidth]{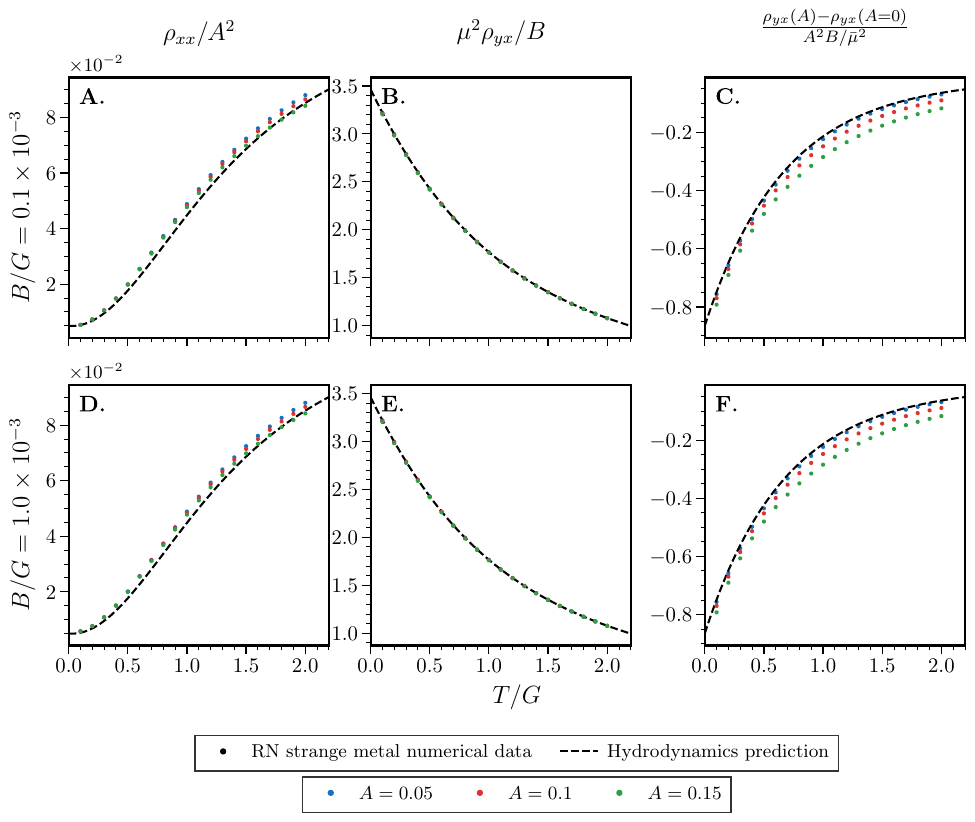}
  \caption{\footnotesize{Comparison of the prediction
      with numerical results from a 2D holographic \acl{rn} strange
      metal model in a periodic potential $\mu_\text{ext}=\bmu + \frac{\bmu
      A}{2}\Bigl(\cos(G x) +A \cos(G y)\Bigr)$ with strength $A$ and two
      different
      perpendicular magnetic field values. The fixed parameter is $G/\bmu =
      \num{0.1}$.
      The longitudinal resistivity $\rho_{xx}$ is compared to the relaxation
      rate prediction $\Gamma_{(2)}/\omega_{p,\text{can}}^2$ in Table
      \ref{tab:summary}; to this numerical accuracy $\Gamma_{(4)}$ is
      negligible.
      The Hall resistivity is compared to the prediction
      $\rho_{yx}=\frac{B}{\bar{n}}$, Eq.~\eqref{eq:full-resist}, which should
      hold
      for low $A$.
      The
      third column
      extracts the subleading order in the Hall resistivity
      $\rho_{yx}(A)-\rho_{yx}(A=0)$. For small $A \sim 0.05$ the subleading
      contribution to $\rho_{yx}$ is the non-dissipative hydrostatic correction
      $\bar{n} - \barv{n}{0} = \barv{n}{2}$ in
      Eq.~\eqref{eq:higher-order-suscep}.
      For larger values of $A$ we see deviations from the perturbative
      prediction.
  }}
  \label{fig:weak-hydro-extraction}
\end{figure}

Fig.~\ref{fig:weak-hydro-extraction} shows the comparison of the numerical
results with weak-lattice hydrodynamics for two different values of the
magnetic field. For low $A$, hydrodynamics is an excellent match.
Specifically, it
already directly tests that $\rho_{yx} = B/\bar{n} + \ldots$ to first
subleading order in that there is no additional correction of size $A^2B$ {\em
aside} from the induced change in overall total density
$\bar{n}=\barv{n}{(0)}+ \barv{n}{2}+\ldots$. In the set-up here where the
lattice is imprinted through the chemical potential, $\barv{n}{2}$ simply
equals the expression in Eq.~\eqref{eq:higher-order-suscep}.
Fig.~\ref{fig:weak-hydro-extraction} (middle column) shows a perfect agreement,
including this change in $\bar{n}$. How well it matches can be seen in
Fig.~\ref{fig:weak-hydro-extraction} (right) displaying the comparison
\mbox{$\frac{\bmu^2}{A^2B}(\rho_{yx}-\rho_{yx}(A=0)) = -
  \frac{\bmu^2}{2\barv{n}{0}^2}\frac{\partial \barv{\cnn}{0}}{\partial
\mu}\Big\rvert_{A=0}|\hat{\mu}(\bk)|^2$}.

At leading order, $\rho_{xx} =
\frac{\Gamma_{(2)}}{\omega_{p,\text{can}}^2}+\ldots$ already matches the
hydrodynamically derived expression exactly in agreement with
\cite{balmTlinearResistivityOptical2023,chagnetHydrodynamicsRelativisticCharged2023}.
The next order corrections of order $A^4 \sim 10^{-8}, \frac{B^2}{\bmu^4} \sim
10^{-6}$ are not extractable at this time due to the increase in numerical
error at low $T/G$ (the size of the numerical error is shown in
Fig.~\ref{fig:influence-grid-points}).%

\begin{figure}
  \centering
  \includegraphics[width=0.9\textwidth]{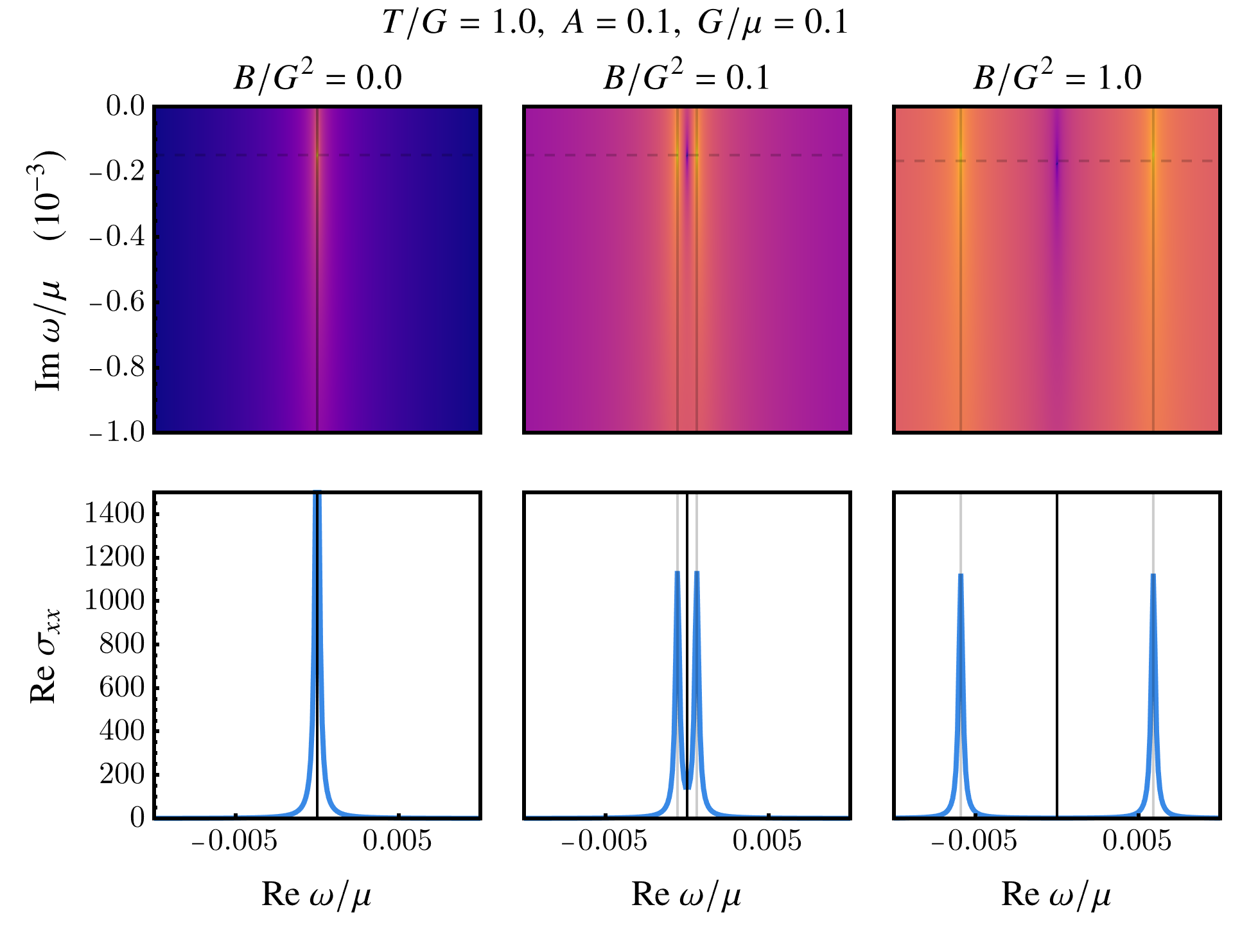}
  \caption{\footnotesize{The AC longitudinal conductivity $\sigma_{xx}(\omega)$
      computed from a numerical hydrodynamics computation in the presence of
      the 2D
      lattice $\mu_\text{ext}=\bmu + \frac{\bmu A}{2}\Bigl(\cos(G x) +A \cos(G
      y)\Bigr)$ and magnetic field for complex frequencies
      $\omega=\omega_{R}+i\omega_I$. The top gives a density plot of the
      absolute
      value $|\sigma_{xx}|=\sqrt{(\realPart \sigma_{xx})^2+ (\imagPart
      \sigma_{xx})^2}$; the bottom only the real part on the real frequency
      axis.
      The two cyclotron poles are clearly visible. Their location on the
      imaginary
      frequency axis is denoted by the horizontal dotted line in the density
      plots:
      this denotes the leading order momentum relaxation rate $\tau_0^{-1} +
      \sigma_Q B_0^2/\chi_{\pi\pi}$. The vertical dotted lines in both plots
      indicate the expected position of the cyclotron pole
      \eqref{eq:final-cyclotron-drude}. Thermodynamic data and transport
      coefficients are taken from the \acl{rn} strange metal model
      equation of
      state Eq.~\eqref{eq:Gibbs-pot}.
  }}
  \label{fig:position_poles_hydro}
\end{figure}

\begin{figure}
  \centering
  \includegraphics[width=0.9\textwidth]{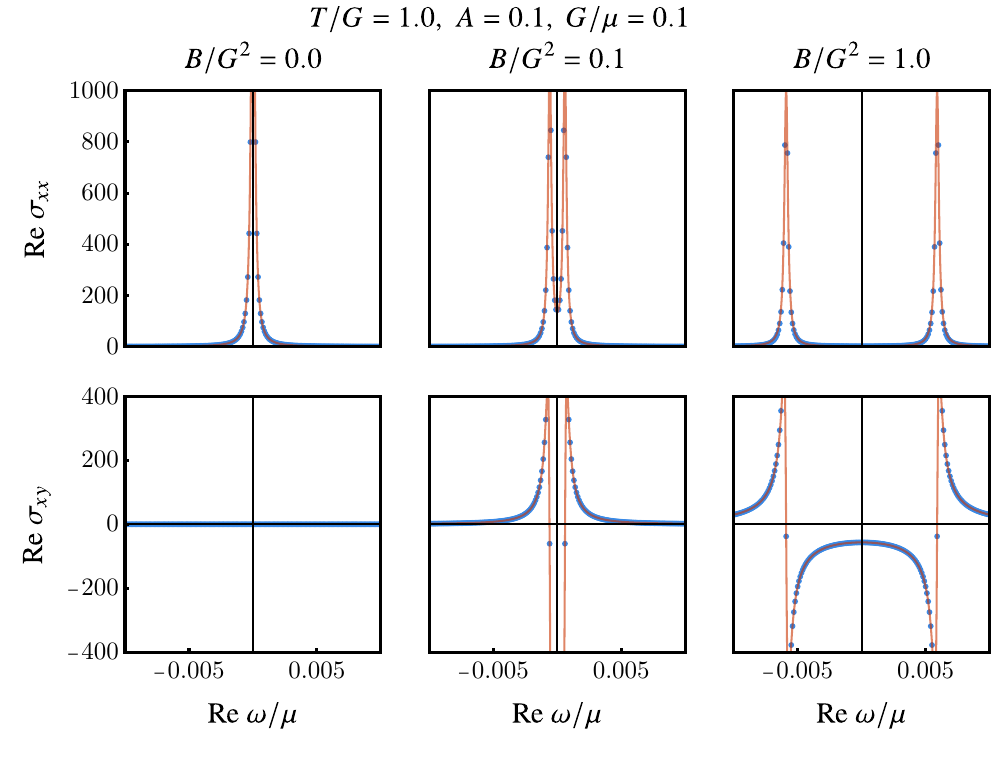}
  \caption{\footnotesize{Comparison of numerically computed AC hydrodynamics
      conductivities with the analytical prediction for weak lattice
      hydrodynamics
      Eq.~\eqref{eq:true-cond}. Thermodynamic data and transport coefficients
      are
      taken from the \acl{rn} strange metal model equation of state
  Eq.~\eqref{eq:Gibbs-pot}.}}
  \label{fig:comparison-hydro-drude}
\end{figure}

To test the more important predictions: the cyclotron frequency shift, the
Drude weight shift and the relation $\rho_{yx} =
\frac{\omega_c}{\omega_p^2}+\ldots$ we need to combine this with the AC
computation of pure hydrodynamics in the presence of a background periodic
modulation of the chemical potential. In such a computation we can extract
with great precision the location of the two leading hydrodynamic poles as a
function of temperature and lattice strength. These form an experimental
measurement of the true cyclotron frequency (the real part of pole) and
momentum relaxation rate (the imaginary part) in the hydrodynamics regime.
Fig.~\ref{fig:position_poles_hydro} shows this for one specific value of the
temperature and lattice strength. As expected, these hydrodynamics numerical
computations showcase the Drude weak translational symmetry breaking physics
in the presence of a magnetic field we have described so far in this paper
encoded in the analytic conductivity expression Eq.~\eqref{eq:true-cond}.
Fig.~\ref{fig:comparison-hydro-drude} illustrates this.

\begin{figure}
  \centering
  \includegraphics[width=\textwidth]{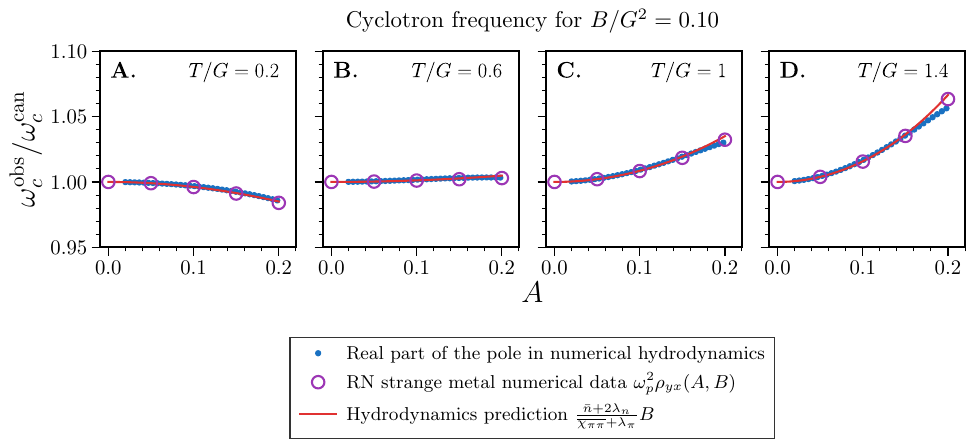}
  \caption{\footnotesize{The cyclotron frequency shift expressed $\omegacobs$
      relative to the canonical cyclotron frequency
      $\omegaccan=\frac{\barv{n}{0}
      B}{\barv{\chipp}{0}}$ from a full AC numerical hydrodynamics calculation
      (blue
      dots) with the black hole transverse transport data ($\omega_p^2$ is
        computed
      using \eqref{eq:drude-weight-correction}; purple circles) and the
      hydrodynamics analytical formula \eqref{eq:final-cyclotron-drude}  (red
      line) .
  }}
  \label{fig:comparison_pole_hydro_real}
\end{figure}

\begin{figure}
  \centering
  \includegraphics[width=\textwidth]{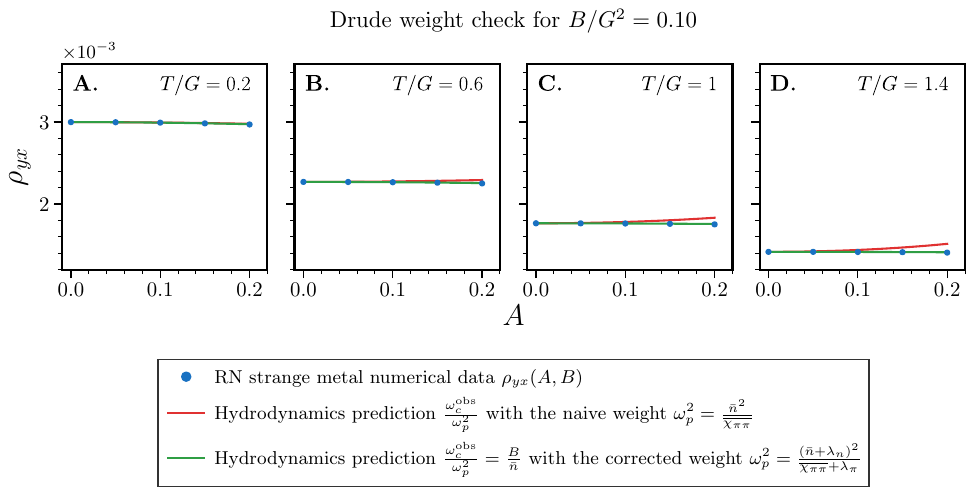}
  \caption{\footnotesize{Comparison of the transverse resistivity $\rho_{yx}$
      from the black hole transport data (blue) with the hydrodynamics
      prediction
      $\rho_{yx} = \frac{\omegacobs}{\omega_p^2}$ where $\omega_p^2$ is computed
      using \eqref{eq:drude-weight-correction} (green line) and using the
      canonical
      formula $\omega_p^2 = \frac{\barv{n}{0}^2}{\barv{\chipp}{0}}$ (red line).
  }}
  \label{fig:comparison_hydro_weight}
\end{figure}

\begin{figure}
  \centering
  \includegraphics[width=\textwidth]{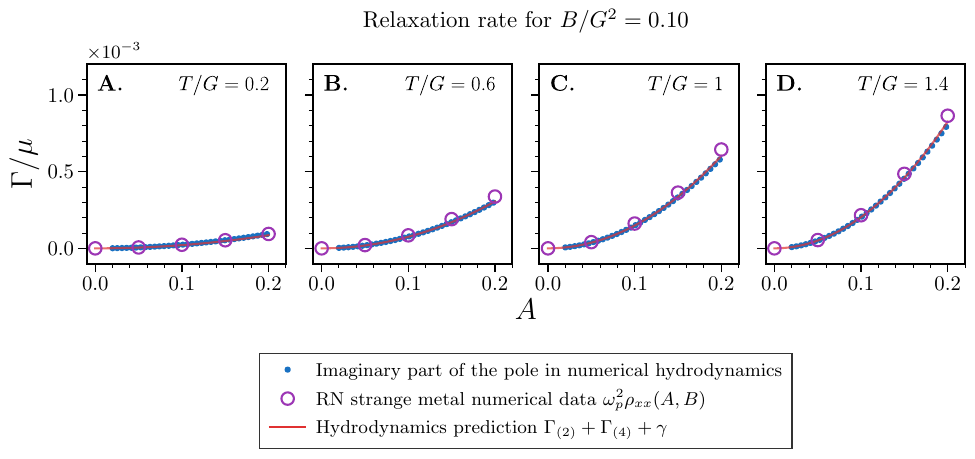}
  \caption{\footnotesize{Comparison of the imaginary part of the pole from a
      full numerical hydrodynamics calculation (blue dots) with the black hole
      longitudinal transport data ($\omega_p^2$ is computed using
      \eqref{eq:drude-weight-correction}; purple circles) and the hydrodynamics
      analytical formula from Table~\ref{tab:summary} (red line).
  }}
  \label{fig:comparison_pole_hydro_im}
\end{figure}

We can then compare these values to the various analytical formulae we have
derived and to the black hole transport data as we vary the lattice strength
$A$ and the temperature $T/G$ to exhibit the remaining predictions from weak
lattice hydrodynamics. Fig.~\ref{fig:comparison_pole_hydro_real} shows
precisely the predicted cyclotron frequency shift $\omega_A$ for low $A$ in
numerical AC hydrodynamics as predicted from our analytical computation. It
{\em also} shows the validity of the expression
$\rho_{yx}=\omegacobs/\omega_p^2$ by using the computed Hall resistivity from
the strange metal model holographically dual to a \ac{rn} black hole. An important aspect
here is that we used the order $A^2$ corrected $\omega_p^2$ to obtain
$\omegacobs = \omega_p^2\rho_{yx}$. Fig.~\ref{fig:comparison_hydro_weight}
shows the \ac{rn} strange metal model numerical data does not match with the Hall
conductivity if only the corrected cyclotron frequency is used, but not the
corrected Drude weight. At the  order we can compute numerically, the
imaginary part of the cyclotron pole extracted from $\Gamma =
\omega_p^2\rho_{xx}$ also matches the hydrodynamic prediction; see
Fig.~\ref{fig:comparison_pole_hydro_im}.

\begin{figure}
  \centering
  \includegraphics[width=0.9\textwidth]{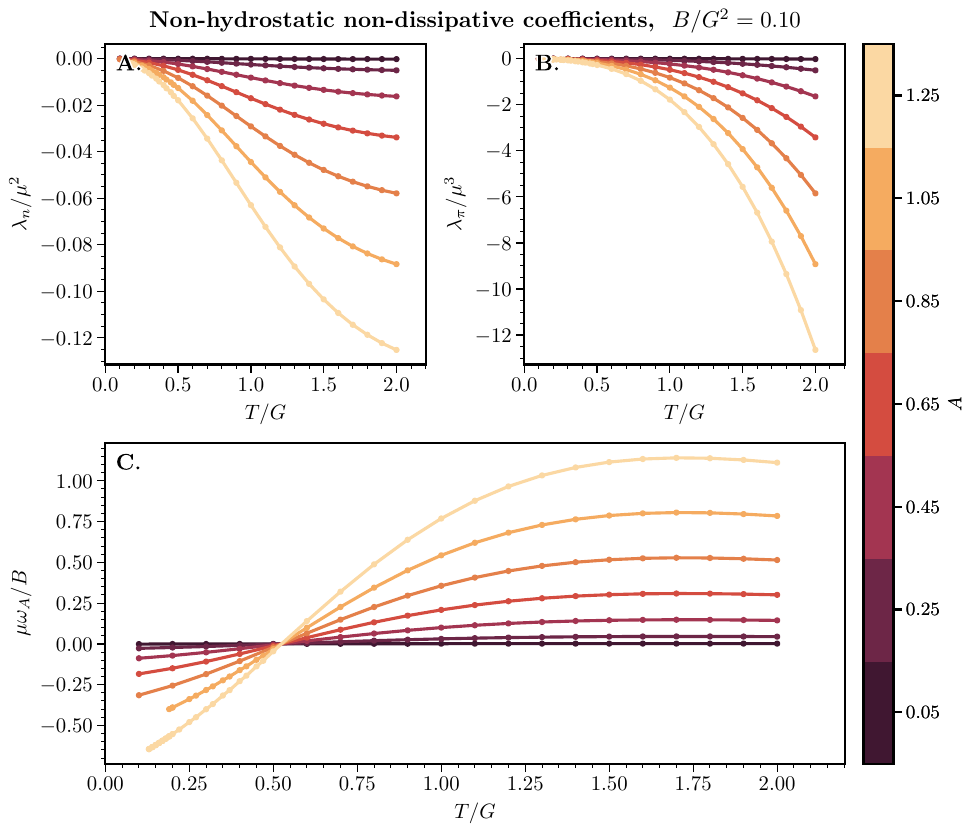}
  \caption{\footnotesize{Non-hydrostatic corrections from
      Table~\ref{tab:summary} evaluated on the black hole data for various $A$,
      $T/G$ at fixed $B/G^2 = \num{0.1}$ and $G/\bar \mu^2 = \num{0.1}$.
  }}
  \label{fig:non-hydrostatic-corrections}
\end{figure}

In much of the weak lattice regime these corrections are small, but they are
definitely there. As explained, the hydrostatic corrections
$\bar{n}=\barv{n}{0}+\barv{n}{2}+\ldots $ are not unexpected, but the
non-hydrostatic corrections are and are new. Using their explicit expressions
in Table~\ref{tab:summary} we can show their size using the thermodynamic
values and transport coefficients extracted from the numerical \ac{rn} strange
metal model simulations:
Fig.~\ref{fig:non-hydrostatic-corrections}. One notices that they become
rapidly important as the lattice strength $A$ increases, especially
$\lambda_\pi$. The relevant comparison scale is $\barv{n}{2} =
\frac{\bar{\mu}^2A^2}{8} \frac{\partial^2 n}{\partial \mu^2} =
\frac{\bar{\mu}^2A^2}{8 \sqrt{3}}$ for the external chemical modulation
Eq.~\eqref{eq:ionic-potential-cosine} in the \ac{rn} model.

\subsection{Hydrodynamics predictions in the Gubser-Rocha strange metal model}
\label{sec:gubser-rocha}

We have shown that the small lattice regime of strange metals dual to black
holes is dominated by universal hydrodynamics behaviour. The first five of our
observational conclusions were shown to be evident in the numerical
simulations: the shift in the cyclotron frequency, the persistence of the
relation $\rho_{yx}=\omega_c/\omega_p^2$, the persistence of the relation
$\rho_{yx}=B/\bar{n}$; the contribution of $\gamma$ to $\rho_{xx}$ and
implicitly therefore a Hall angle determined by these effects. Implicitly we
have also shown the universality of the results by using thermodynamic
quantities and appropriate transport coefficients in our hydrodynamical
expression to obtain the matching values.
Besides serving as evidence of the fundamental role of hydrodynamics in
strongly correlated systems, this observation also invites us to use our
computation for a prediction for what the small lattice regime predicts in a
strange metal that is closer to those observed experimentally in high $T_c$
cuprates. The prime candidate for this is the so-called \acl{gr}
black hole. The main difference between the \ac{gr} and \ac{rn} black holes lies in the
addition of a marginal scalar operator which deforms the theory  while
preserving the local quantum critical nature of the underlying physics
\cite{gubserPeculiarPropertiesCharged2010,chagnetProperQuantizationGubserRocha2022}.
This quantum critical IR has has a linear-in-$T$ resistivity $\rho \sim T$ and
Sommerfeld entropy $s \sim T$ at small $T$ similar to that observed in high
$T_c$ cuprates
\cite{niuDiffusionButterflyVelocity2017,Jeong:2018tua,Smit:2021dwh,balmTlinearResistivityOptical2023,ahnHolographicGubserRochaModel2023}

To predict the results of weak-lattice hydrodynamics for the \ac{gr} strange metal model,
we therefore only change to the appropriate equation of state $P(T,\mu)$ and
transport coefficients $\eta,\zeta, \sigma_Q$ from their \acl{rn}
model values to the appropriate \acl{gr} values
\cite{gubserPeculiarPropertiesCharged2010,chagnetHydrodynamicsRelativisticCharged2023}.\footnote{We
  make here the assumption that the choice of quantization of the boundary
  scalar operator is such that UV conformal symmetry is preserved— the dilaton
  must be a marginal deformation. One must therefore use mixed boundary
  conditions for the dilaton at the boundary
\cite{chagnetProperQuantizationGubserRocha2022}.}
\begin{align}
  \label{eq:Gibbs-pot}
  \hspace*{-.2in}
  \begin{array}{*2{>{\displaystyle}l}}
    P_{\text{RN}}(\mu,T) = \frac{\mu^4}{2}\left(\frac{\mu^2+8\pi^2T^2-2\pi
    T\sqrt{3\mu^2+16\pi^2T^2}}{(\sqrt{3\mu^2+16\pi^2T^2}-4\pi T)^3}\right)~,
    & P_{\text{GR}}(\mu,T) =\frac{(3\mu^2+16\pi^2 T^2)^{3/2}}{27}  \\
    \eta_{\text{RN}} = \frac{s_{\text{RN}}}{4\pi} =
    \frac{1}{4\pi}\left(\frac{\partial P}{\partial T}\right)_{\mu} ~,
    & \eta_{\text{GR}} = \frac{s_{\text{GR}}}{4\pi} =
    \frac{1}{4\pi}\left(\frac{\partial P}{\partial T}\right)_{\mu} \\
    \zeta_{\text{RN}} = 0 ~,
    & \zeta_{\text{GR}}=0 \\
    \sigma_{Q,\text{RN}} = \frac{4\pi^2
    T^2}{9}\left(\frac{\sqrt{3\mu^2+16\pi^2T^2}-4\pi T}{\mu^2+8\pi^2T^2-2\pi
    T\sqrt{3\mu^2+16\pi^2T^2}}\right)^2 ~,
    & \sigma_{Q,\text{GR}}
    =\left(\frac{3\mu^2+16\pi^2T^2}{16\pi^2T^2}\right)^{-3/2}
  \end{array}
\end{align}
The expression $\eta = \frac{s}{4\pi}$ is not universal, but uses one of the
findings that holographic strange metal models always have such a minimal viscosity
set by the entropy density. From this we can deduce the cyclotron frequency
shift using the expression in Eq.~\eqref{eq:omega-c-shift}:
\begin{align}
  \frac{\omegacobs}{\omegaccan} & \simeq  1 - \frac{A^2}{2}
  \left(1-\frac{G^2}{6\mu^2}\right)  + \frac{2 A^2 \pi}{\sqrt{3}} \frac{T}{\mu}
  + \frac{A^2 \pi^2}{9}\biggl( \frac{T}{G} \biggr)^2 + \ldots \quad &  &
  \text{(RN)} \nonumber \\
  \frac{\omegacobs}{\omegaccan} & \simeq  1 + \frac{A^2}{2} \left(\frac{1}{12}
  -\frac{G^2}{\mu^2}\right) + A^2 \Bigl(3 + \frac{G^2}{\mu^2} \Bigr) \Bigl(
  \frac{4 \pi T}{3 \mu} \Bigr)^2 + \ldots \quad                        &  &
  \text{(GR)}
  \label{eq:GR-prediction-cyclotron}
\end{align}
There is a noteworthy aspect in these expressions: In the hydrodynamic regime
where they are valid both the lattice and the lattice strength  must be small:
$G\ll \mu$ and $A\ll 1$. That means that for the \ac{rn} model for all ${G}\ll\mu$
and for the \ac{gr} model for $\frac{1}{\sqrt{12}} \ll G/\mu \ll 1$ at low $T/G$
the actual value of $\omegacobs$ decreases somewhat with respect to the
canonical value. This can be seen in Fig.~\ref{fig:comparison_pole_hydro_real}.
This can be interpreted as an early signal of possible new behaviour where the
subleading correction becomes important. Naively extrapolating to a value of
$A^2\geq 2$ such that the first two temperature-independent terms nearly
cancel, the subleading temperature dependence would in fact take over. This is
of course outside of the hydrodynamic regime where these expressions are valid,
but it could indicate the onset of some novel apparent scaling behaviour for
larger $A$ as we also see in our numerical simulations.

\section{Hydrodynamical magnetotransport in experiment and
  strong lattice effects as a possible explanation of the Hall angle anomaly in
the cuprate strange metals.}
\label{sec:strong-lattice-and-cuprate-hall}

Let us first make a few more general points on the hydrodynamic effects of
magnetotransport with weak momentum relaxation %
combined with their qualitative observational consequences.
\begin{itemize}
  \item[1.]
    The most direct is continued cross-consistency
    $\rho_{yx}=\frac{\omegacobs}{\omega_p^2}$ and the extended range of validity
    of the Hall coefficient as  measurement of the effective charge density
    $R_{H}=\frac{1}{\bar{n}}$ despite the shifted cyclotron frequency
    $\omegacobs\neq \frac{\bar{n}B}{\overline{\chipp}}$.
    The cyclotron frequency and the Drude weight $\omega_p^2$ can be directly
    measured from the optical conductivity in finite magnetic field in the
    regime
    where there is a Drude-like peak as in
    \cite{postObservationCyclotronResonance2021,legrosEvolutionCyclotronMass2022}.
    These can be compared with precision transport measurements of the Hall
    resistivity $\rho_{yx}$ as in \emph{e.g.},
    \cite{badouxChangeCarrierDensity2016,putzkeReducedHallCarrier2021,ayresIncoherentTransportStrange2021}
    to establish whether the relation holds or not. If it does, this is a
    possible
    indication of weak/medium strength translational symmetry breaking
    hydrodynamics.

  \item[2.]
    This cyclotron shift cannot be reabsorbed in a ``renormalization'' of the
    effective mass ${\chi_{\pi\pi}}=\bar{\eps}+\bar{P}$ ($={nm_{\star}}$ in a
    Fermi Liquid). When the cyclotron frequency is written as an effective
    cyclotron mass $m_c= B/\omega_c$ (in units where the elementary charge
    $e=1$),
    this is in fact what is meant. Here this is evident not only in the
    non-hydrostatic corrections $\lambda_i$, but also in that these
    non-hydrostatic corrections differ by a factor 2 in the cyclotron frequency
    and the plasmon frequency/Drude weight $\omega_p^2$ for which we now have
    $\omega_p^2 \neq \chi_{\pi j}^2/\chi_{\pi\pi} \neq \chi_{\pi j}\omega_c/B$.
    But
    in a hydrodynamic regime this non-renormalizability is even more obvious in
    the more general sense that there is no direct relation with the the
    specific
    heat 
    $c_V = T\chi_{ss} =\frac{1}{T}\left(\chi_{\eps \eps} - 2\mu \chi_{\eps n}+\mu^2\chi_{nn}\right)$.
    This only works if one has a
    microscopic interpretation in terms of quasiparticles, e.g, for a Fermi
    Liquid
    where (in 2D) $c_V = C \cdot m_{\star} T$ and $\omega_p^2 = n/m_{\star} = C
    n^2 T/c_V \equiv C n^2/\gamma_{c_V}$.\footnote{This is the insight behind
      the
      Kadowaki-Woods ratio $R_{KW} =\rho / (c_V^2) =
      \frac{\omega_p^2}{\gamma_{c_V}^2} \frac{1}{T^2}\frac{1}{\tau} + \ldots$
      which
      is constant for a 2D Fermi-liquid $\tau\sim 1/T^2$ at low $T$, but not so
      for
      a cuprate strange metal \cite{husseyNongeneralityKadowakiWoodsRatio2005}.
      See,
      however, \cite{legrosUniversalTlinearResistivity2019}, which studies the
      ratio
      $R_{LT}=\rho/c_V =\frac{\omega_p^2}{\gamma_{c_V}}\frac{1}{T\tau}$ that
      would
    be more natural for strange metals with $T$-linear resistivity.}

    A recent far more phenomenological approach to the optical conductivity
    \cite{michonPlanckianBehaviorCuprate2022a} argued that optical mass
    enhancement $m_{\text{opt}}(\omega)$ defined from the parametrization
    \begin{align}
      \sigma_{xx}(\omega) = \frac{K}{1/\tau(\omega)-i\omega
      m_{\text{opt}}(\omega)/m}
    \end{align}
    in cuprates could be reconciled with the specific heat assuming that
    $c_v\simeq C m_{\text{opt}}(0;T) T$. Moreover, the SYK lattice model in
    \cite{patelUniversalTheoryStrange2022,liStrangeMetalSuperconductor2024}
    exhibits this phenomenology. In this approach/these models the effective
    Drude
    weight is $\omega_{p,\text{eff}}^2 = K/m_{\text{opt}}(0;T)$ suggesting
    precisely an effective mass renormalization correlated with the specific
    heat
    that one would only expect in a quasiparticle-like theory, even though SYK
    models do not have long-lived quasiparticles in which transport can be
    understood. Though the phenomenological approach to the conductivity is
    based
    on memory functions and completely consistent with hydrodynamics at low
    $\omega/T$ \cite{gotzeHomogeneousDynamicalConductivity1972}, the persistence
    of model-specific cross-consistency with the specific heat is surprising in
    this non-quasiparticle context. An extension of the memory function
    technique
    to the combined thermo-electric transport could elucidate the underlying
    reason.

  \item[3.] This
    inability to absorb the cyclotron shift in a ``renormalized'' mass is
    qualitatively analogous to the disconnect in ordinary Fermi liquids with an
    anisotropic and non-quadratic dispersion relation between the band mass and
    the cyclotron mass inferred from the cyclotron frequency. The latter in that
    case is qualitatively an ``averaging'' over the band mass (see \emph{e.g.},
    \cite{singleton2001band}), quite analogous to how the
    hydrostatic part of the cyclotron frequency shift arises here from an
    ``averaging'' over the ratio $\overline{\frac{\langle
    \chi_{\pi j}\rangle}{\langle\chi_{\pi\pi}\rangle}}=\overline{\frac{n}{\eps+P}}$
    --- though there is no analogue of the non-hydrostatic corrections
    $\lambda_i$. However, in a Fermi-liquid or in any 2D quasiparticle theory
    the
    cyclotron frequency in a transverse magnetic field is always directly
    proportional to the density of states at the Fermi level
    that also sets the specific heat (see \emph{e.g.},
    \cite{tamuraAnalysisHaasvanAlphen1994,merinoCyclotronEffectiveMasses2000}.),
    whereas that is generically not the case in the hydrodynamic regime as
    discussed
    at length above.

\end{itemize}
These hydrodynamical insights may be relevant to the surprising experimental
finding in
\cite{legrosEvolutionCyclotronMass2022} that in high $T_c$ cuprates the
cyclotron frequency (the inverse of the cyclotron mass) decreases with doping.
As in the ionic lattice potential we use to break translational symmetry
$\mu(x,y) = \bmu + \frac{\bmu A}{2}\Bigl(\cos(Gx)+\cos(Gy)\Bigr)$ an increase
in the lattice strength goes hand-in-hand with an increase in density
$\bar{n}=\barv{n}{0}+\barv{n}{2}+\ldots$ there is an argument to be made that
an increase in $A$ effectively changes the doping of our model as well.
Assuming this, the change in the cyclotron frequency in the first figure in
Fig.~\ref{fig:comparison_pole_hydro_real} at $T/G=0.2 $ as a function of $A$
has the right trend in the change in cyclotron mass as a function of doping
(Fig.~3 in \cite{legrosEvolutionCyclotronMass2022}) or as a function of
disorder strength in the SYK study (Fig.~11 in
\cite{guoCyclotronResonanceQuantum2023}).

Note from Fig.~\ref{fig:comparison_pole_hydro_real} that this is the
temperature regime where the anomalous shift $\omega_A$ has the opposite sign
as $\omega_c$. As we discussed in our predictions for the possibly
experimentally more relevant \acl{gr} model, by extrapolating to larger
$A$, it may therefore cause a sign change in the Hall response. Such a sign
change has been observed \cite{ginsbergHallEffectYBa2Cu3O71994}, albeit just
below $T_c$ where superconducting fluctuations, which are not taken into
account here, should also play a role.

The cyclotron change with doping is also compared to the anomalous $p$ to $1+p$
change in the Hall coefficient $R_H$ observed in high $T_c$ cuprates
\cite{andoEvolutionHallCoefficient2004,padillaConstantEffectiveMass2005a,tsukadaNegativeHallCoefficients2006,badouxChangeCarrierDensity2016,putzkeReducedHallCarrier2021,ayresIncoherentTransportStrange2021}
(Fig.~4 in \cite{legrosEvolutionCyclotronMass2022}).
Our results here show that such a direct comparison between the Hall
coefficient and the cyclotron frequency must be done very carefully: the
expression of the cyclotron frequency in terms of the charge density $\bar{n}$
is corrected, whereas the Hall coefficient is not to first subleading order.
This point has also recently been made in
\cite{gouterauxDrudeTransportHydrodynamic2023}.

\subsection{Strong lattice potentials and a second time scale.}

Despite this tempting comparison between our perturbative lattice
magneto-hydrodynamical results and cyclotron experiments in the cuprates, it
is also obvious from these general observational consequences that weak
translational symmetry breaking hydrodynamics
cannot provide full explanation of the observed anomalous Hall angle scaling in
the cuprate strange metals.
Not only are all quantities given by only a small correction to their dominant
Drude single momentum relaxation response, in this particular framework the
Hall resistivity is also unchanged due the remarkable identity below
Eq.~\eqref{eq:full-resist}. Specifically
in charged hydrodynamics with weak translational symmetry breaking
\begin{align}
  \cot\thetaHall & = \frac{\rho_{xx}}{\rho_{yx}} =
  \frac{A^2}{\omega_p^2\tau_0}\frac{\bar{n}^{(0)}}{B} +{\cal O}(A^4B, B^3)
  \non
  \sigma_{xx}    & = \omega_p^2 \frac{\tau_0}{A^2} + {\cal O}(A^4)
\end{align}
Only for large lattice strengths $A\geq 1$ where the perturbative series gets
re-summed, can a different scaling regime emerge where $\cot(\thetaHall)\sim
T^{n}$ behaves as a single power different from the power of the longitudinal
resistivity over a large temperature range as observed in
experiment.\footnote{This latter statement, that one must go beyond a weak
  lattice approximation, is even true, if one takes a more general
  phenomenological description where one simply adds a cyclotron shift by hand
  in the Drude model, rather than from a more coherent underlying theory.
  Such a shift in the cyclotron frequency
  can be (incorrectly) interpreted as a second dissipative timescale, if one
  only measures the DC longitudinal and Hall conductivity.
  \begin{align}
    \sigma_{ij} = \omega_p^2 \frac{\tau}{1+((\omega_c+\omega_A)\tau)^2}
    \begin{pmatrix}
      1                        & (\omega_c +\omega_A)\tau \\
      -(\omega_c+\omega_A)\tau & 1
    \end{pmatrix}~.
  \end{align}
  An artificial redefinition of $\omega_c+\omega_A= \omega_c
  \tau_{\text{Hall}}/\tau$
  \begin{align}
    \label{eq:magneto-hydro-redefinition-cyclotron}
    \sigma_{ij} = \omega_p^2 \frac{\tau}{1+(\omega_c\tauHall)^2}
    \begin{pmatrix}
      1                 & \omega_c\tauHall \\
      -\omega_c\tauHall & 1
    \end{pmatrix}~,
  \end{align}
  makes it appear as if there is a second dissipative timescale $\tauHall$
  instead.
  In particular Eq.~\eqref{eq:magneto-hydro-redefinition-cyclotron} implies
  $\cot(\thetaHall) \sim \omega_c\tauHall$, $\sigma_{xx} \sim \omega_p^2\tau $.
  The converse to this, that a second time-scale can also be interpreted as a
  cyclotron frequency shift, was already pointed out long ago
  \cite{romeroCyclotronResonanceCuprate1992}.
  It behooves emphasizing that
  the correct
  redefinition/interpretation
  rests on whether there is a truly second %
  {dissipative relaxational} timescale.
  If one measures DC transport alone one cannot distinguish
  these, as also emphasized in \cite{graysonSpectralMeasurementHall2002a}.
  Relevant for the argument in this article is that weak translational symmetry
  breaking will always give Drude-like answers.
}

These discussions point to a clear conclusion: to explain the Hall angle
anomaly and other peculiar magnetotransport characteristics in cuprate strange
metals strong electron-electron correlations and absence of quasiparticles are
not enough. A system with strong electron-electron correlations and absence of
quasiparticles will have transport governed by hydrodynamics provided the
translational symmetry breaking scale is weak and at a scale larger than the
electron mean free path, but this does not have the phenomenology of cuprate
magnetotransport. Models for magnetotransport in real cuprate strange metals
must therefore either have strong translational symmetry breaking such as
\cite{patelMagnetotransportModelDisordered2018}, although this is difficult to
square with the observed Drude-like response in the optical conductivity at
low $T$, and/or an electron mean free path that is of the same order as the
translational symmetry breaking scale, in which case the hydrodynamic analysis
employed here does not apply.

\begin{figure}
  \centering
  \includegraphics[width=0.9\textwidth]{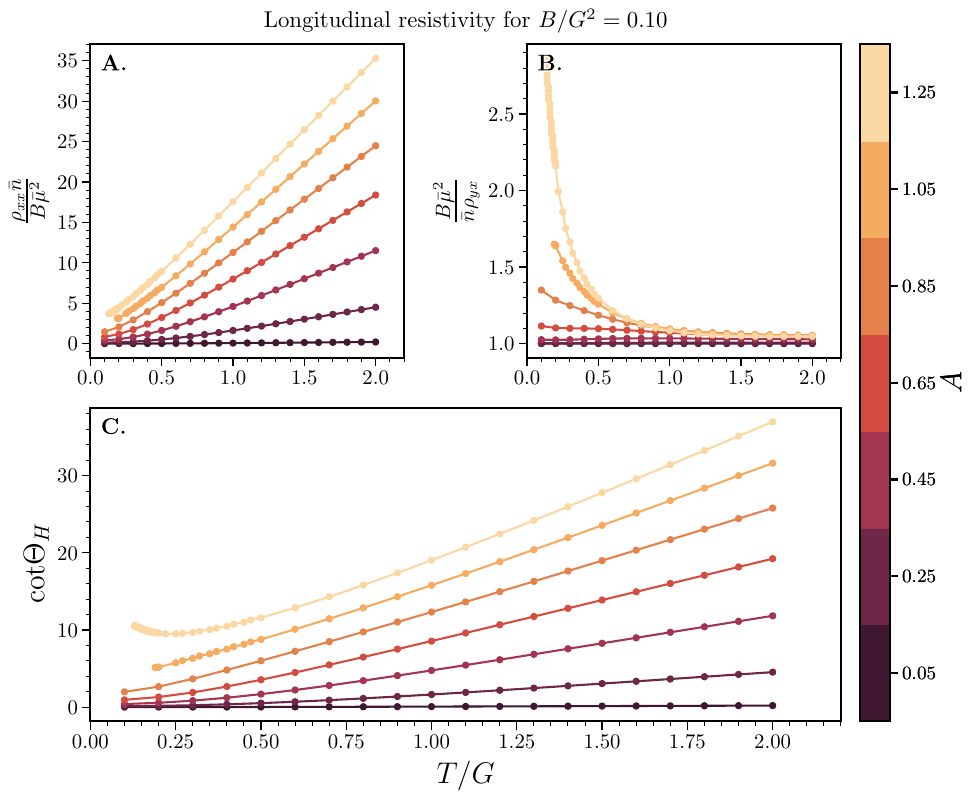}
  \caption{\footnotesize{Decomposition of the cotangent of the Hall angle into
      two pieces $\cot(\thetaHall) = \frac{\rho_{xx} \bar n}{B} \times
      \frac{B}{\bar
      n \rho_{yx}}$. At low values of the lattice strength $A$ where
      hydrodynamics
      dominate the response, the second piece is nearly one and the Hall angle
      is
      entirely constrained by $\rho_{xx}$. As $A$ is increased, this simple
      logic
      ceases to function and the transverse resistivity dominates the low
      temperature scaling behaviour. All quantities are normalized by
      appropriate
      factors of $\bmu$.
  }}
  \label{fig:hall-coefficient-and-rhoxx}
\end{figure}

Both SYK-like models with a microscopic Hamiltonian and holographic methods
with a UV completion can in principle probe beyond the hydrodynamic regime.
We can test this large $A$ regime here numerically in the \ac{rn} model as already
shown in Fig.~\ref{fig:DCCondVsT}. Indeed one sees a different  temperature
dependence between the longitudinal conductivity and Hall angle is observed.
We can exhibit in more detail that entirely novel transport physics is at play
here.
We can decompose the cotangent of the Hall angle as $\cot(\thetaHall) =
\frac{\rho_{xx} \bar n}{B} \times \frac{B}{\bar n \rho_{yx}}$ such that the
second term is trivially equal to one in the regime of hydrodynamics where the
temperature dependence of $\cot(\thetaHall)$ is entirely constrained by that
of $\rho_{xx}$. We plot this decomposition in
Fig.~\ref{fig:hall-coefficient-and-rhoxx} and we see that, for intermediate
values of the lattice strength, not only is the transverse resistivity not
simply given by the average charge density anymore, but it acquires a
temperature dependence which dominates the low-temperature regime of the Hall
angle. This is a promising indication that strong lattice/strong momentum
relaxation effects could be the explanation behind the observed Hall angle
conundrum in the cuprates, although again in such a bad/incoherent metal
scenario one would not expect a Drude peak as is observed below $T\sim
\SI{200}{\kelvin}$ in experiment or in similar numerical simulations for the
\ac{rn} model \cite{balmTlinearResistivityOptical2023}.  Taken together this
suggests that the construction of a theoretical framework to understand
magnetotransport and Hall responses of an incoherent metal in the presence of
strong translational symmetry breaking as an extension of the longitudinal
thermoelectric response \cite{hartnollTheoryUniversalIncoherent2015} is the
next clear question to be answered. In such incoherent systems in an external magnetic the equilibrium state need not be rotationally or parity invariant, and this is known to have novel Hall-type effects as a consequence at large magnetic field \cite{amorettiMagnetothermalTransportImplies2020,amorettiHydrodynamicsDimensionalStrongly2022b}.
Such a theoretical framework should also address the other glaring
magnetotransport puzzle in the cuprates: the non-analytic quadrature scaling
form of the longitudinal magnetoresistance at large magnetic field $\rho_{xx}
\sim T\sqrt{1+B^2/T^2}$
\cite{hayesMagnetoresistanceQuantumCritical2016,ayresIncoherentTransportStrange2021}.
This conflicts directly with any continuum long-distance type of analysis
including hydrodynamics, and is so far only understandable in a strong
disorder effective medium random resistor network type analysis
\cite{stroudGeneralizedEffectivemediumApproach1975,ramakrishnanEquivalenceEffectiveMedium2017,patelMagnetotransportModelDisordered2018}.

Despite this clear signal that strong lattice/disorder effects and another
timescale must be in play to explain both the Hall angle and the longitudinal
magnetoresistance, our hydrodynamics in the presence of both a magnetic field and a lattice
results do indicate the following. In the Drude-like regime and at smaller
magnetic fields  the interpretation of the Hall coefficient in the cuprates as
effective density is more reliable than one might have thought. However, one
should be careful in using this measurement of the density to deduce the Drude
weight and the cyclotron frequency through canonical expressions, as there are
non-hydrostatic corrections that can be significant.

\acknowledgments
\noindent
Jan Zaanen was a large part of the early stages of this project. We shall sorely miss him. We thank R. Davison, B. Gout\'{e}raux, N. Hussey for discussions, and
especially A. Krikun and D. Rodriguez-Fernandez, who contributed to early
stages of this work. We also thank J. Aretz, O. Moors, J. Post, K. Grosvenor.

This research was supported in part by the Dutch Research Council (NWO)
project 680-91-116 ({\em Planckian Dissipation and Quantum Thermalisation:
From Black Hole Answers to Strange Metal Questions.}), the FOM/NWO program 167
({\em Strange Metals}), and by the Dutch Research Council/Ministry of
Education.
The numerical computations were carried out on the Dutch national Cartesius and
Snellius national
supercomputing facilities with the support of the SURF Cooperative (project
EINF-468, EINF-2777, EINF-6933) as well as on the ALICE-cluster of Leiden
University. We are grateful for their help.

\appendix
\section{Transport in quantum critical metals from numerical tests on
holographic models with a periodic potential.}
\label{app:numerics-RN}

The \acl{rn} and \acl{gr} model are two members of the family
of AdS-Einstein-Maxwell-Dilaton models that capture the physics of a wide
class of local quantum critical states through holographic duality.
Holographic duality states that certain strongly coupled large-$N$ $d+1$-dim
quantum field theories are mathematically equivalent to
general relativity in spacetimes with a negative cosmological constant that
asymptotically approach anti-de-Sitter space. The generating function of the
QFT equals the on-shell action of the gravity theory whose boundary conditions
are equated with the sources of QFT operators; see
\cite{zaanenHolographicDualityCondensed2015,hartnollHolographicQuantumMatter2018}
for a review.
\begin{align}
  Z_{QFT}(J) = \left.\int \!{\cal D}\phi\,
  e^{iS^{\text{AdS-grav.}}}\right|_{\phi|_{\partial AdS} =J}
\end{align}
The Einstein-Maxwell-Dilaton models have the following gravitational action
\begin{align}
  \label{eq:emd-action-reg}
  S^{\text{AdS-grav, EMD}} = & \frac{1}{2\kappa^2} \int\!
  d^{d+2}x\sqrt{-g}\left[R+\frac{d(d+1)}{L^2} - \frac{Z(\Phi)}{4} F_{\mu\nu}^2 -
  \frac{3}{2}g^{\mu\nu}\partial_{\mu}\Phi\partial_{\nu}\Phi - V(\Phi)\right]
  \non
  & -\oint \! d^{d+1}x \sqrt{-h}(2K + 4 + {}^{(d+1)}R_h)
  \\
  & + \oint \! d^{d+1}x \sqrt{-h} \left[ \frac{3}{2} \Phi^2 + 3 \Phi N^\mu
  \partial_\mu \Phi + \Phi^3\right]~,\nonumber
\end{align}
where $K$ is the extrinsic trace, $h$ is the induced metric on a hypersurface
at the AdS boundary with Ricci scalar ${}^{(d+1)}R_h$ and outward-pointing
unit vector $N^\mu$.

A finite charge density in the QFT sourced by a chemical potential $\mu$ is
imposed by the boundary condition that the electrostatic potential on the
gravity side equals $A_t|_{\partial AdS} = \mu$. The generic solution of the
corresponding Einstein-Maxwell-Dilaton equations on the gravity side is a
charged black hole, whose near horizon physics encodes the emergent
non-trivial quantum critical ground state in the QFT. The RG flow to this
ground state is triggered by the chemical potential itself or a relevant
scalar operator dual to the dilaton $\Phi$.
For the AdS$_4$ \acl{rn} model, which has $Z(\Phi)=1, V(\Phi)=0,
\Phi=0$, it is the former; for the AdS$_4$ \acl{gr} model, which has
$Z(\Phi)=\exp(\Phi), V(\Phi)=\frac{d(d+1)}{L^2}(1-\cosh(\Phi)),
\partial_n\Phi|_{\partial AdS}=\frac{Q}{2}$ (where the parameter $Q$ is
  related to the chemical potential $\mu$ and the event horizon radius $z_h$ by
$\mu = \sqrt{3 Q z_h (1+ Q z_h)}/z_h$), it is the latter. The choice of
boundary terms for the scalar sector, previously detailed in
\cite{chagnetProperQuantizationGubserRocha2022}, ensures that the boundary
theory remains conformal.
Including a finite transverse magnetic field corresponds to a dyonic black hole
with both magnetic and electric charge.

For spatially constant chemical potential and dilaton the solutions can be
found a\-nalytically and the value of the regularized on-shell action
\eqref{eq:emd-action-reg} (analytically continued to Euclidean time) equals
the Gibbs potential density $\Omega(T, \mu) = T
S^{\text{AdS-grav,EMD}}_{\text{Euclidean, on-shell}}$ with
$S^{\text{AdS-grav,EMD}}_{\text{Euclidean, on-shell}} = i S^{\text{AdS-grav,
EMD}}$ such that the Euclidean time $\tau = i t$ has a periodicity $\beta =
1/T$. For the homogeneous solutions, the time integral contributes a simple
factor $T$ while the spatial integrals contribute an overall volume factor
$V$. The functional expression of the pressure $P(\mu,T) = \Omega(\mu, T)/V$
for the \ac{rn} and \ac{gr} model are quoted in Eqs.~\eqref{eq:Gibbs-pot}. The transport
coefficients $\eta, \zeta, \sigma_Q$ follow from Kubo relations, but one of
the important discoveries in holographic duality is that they end up being
encoded in horizon properties of the (translationally invariant homogeneous
and isotropic) black hole and can also be analytically determined
\cite{zaanenHolographicDualityCondensed2015,hartnollHolographicQuantumMatter2018}.

\begin{figure}
  \centering
  \includegraphics[width=\textwidth]{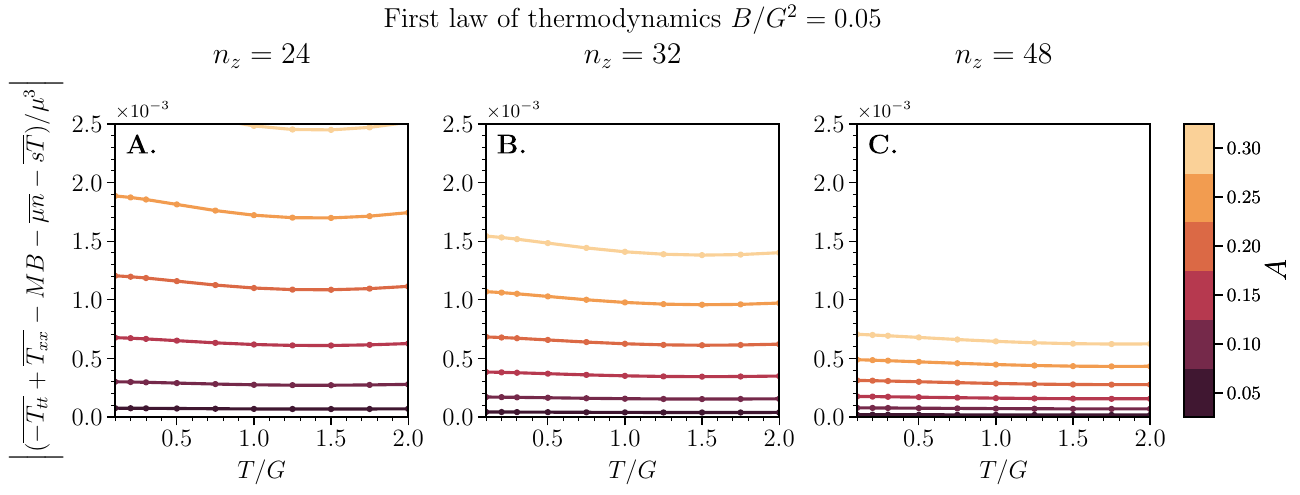}
  \caption{\footnotesize{Numerical tests of the fundamental relation
      $\bar{\eps}+\bar{P}=\bar{s}T+\overline{\mu n}$ of the AdS${}_4$
      \acl{rn} model in the presence of a spatially varying chemical
      potential $\mu(\bx)=\bmu + \frac{\bmu A}{2} \bigl(\cos(Gx)+\cos(Gy)
      \bigr)$.
      The $x,y$-direction are discretized on a $24\times 24$ grid. The barred
      quantities are the spatially averaged ones.
      The deviation observed is clearly a numerical accuracy issue that
      improves when increasing the number of Chebyschev grid points $n_z$ in the
      direction orthogonal to the AdS boundary.
  }}
  \label{fig:fund-rel-test}
\end{figure}

\begin{figure}
  \centering
  \includegraphics[width=\textwidth]{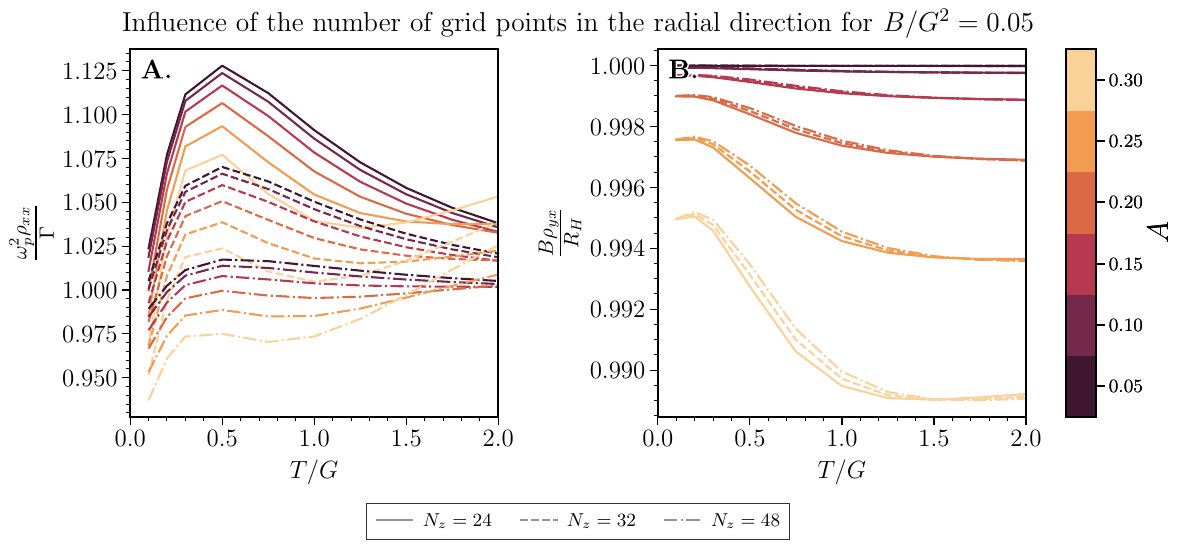}
  \caption{\footnotesize{Influence of the number of grid points in the radial
      direction on the DC resistivities. Each resistivity is normalized by the
      expected hydrodynamics result computed using the analytical formulae we
  derived in this work.}}
  \label{fig:influence-grid-points}
\end{figure}

For a spatially varying solution such as a periodic chemical potential, the
solutions to the AdS-EMD system can only be found numerically. For the
\acl{rn} model at finite magnetic field this is done in the de
Turck-gauge with a Newton-Raphson method on a Chebyschev grid for the finite
distance between the black hole horizon and a cut-off near the boundary of the
spacetime. The code is available at
\cite{floris_balm_2022_7284816,floris_balm_2022_7285050}.
Expressions for thermodynamic quantities $\eps(\bx), n(\bx)$ are extracted from
the appropriately normalized normal derivative to the AdS boundary of the
time-time component of the metric $g_{tt}(\bx,r)$ and the electrostatic
potential $A_t(\bx,r)$ respectively
\cite{zaanenHolographicDualityCondensed2015}. The entropy is the
Bekenstein-Hawking entropy $s(\bx) = \frac{2\pi}{\kappa^2}A_H(\bx)$, %
where $A_H(\bx)$ is the area density of the black hole horizon. In the presence
of a magnetic field $P(\bx)$  equals the pressure $P(\bx)=T_{xx}(\bx)+M(\bx)
B$ with $M$ the magnetization (for a square lattice along the $x,y$-axes the
choice of $T_{xx}$ or $T_{yy}$ is immaterial)
\cite{hartnollTheoryNernstEffect2007,donosDCConductivityMagnetised2016}. To
the precision we work with the magnetization is spatially constant and we can
use its analytic value from the homogeneous \ac{rn} model.
Due to the fact that conserved currents do not renormalize, one can prove that
on the gravity side the DC thermoelectric current response can be read of from
the horizon properties of these solutions, without the need to solve Kubo
relations from linear fluctuations \cite{donosDCConductivityMagnetised2016}.
As always there are trade-offs between accuracy and time. We have used a
$24\times 24$ grid in the $\bx,\by$ directions and a $24$-point Chebyshev grid
in the direction orthogonal to the AdS boundary. As indication of the accuracy
we test the spatial average of the fundamental relation
$\bar{\eps}+\bar{P}=\bar{s}T+\overline{\mu n}$ vs the number of Chebyshev grid
points. The result is in Fig.~\ref{fig:fund-rel-test}. Due to this trade-off
cost, the \acl{gr} model at finite magnetic field and spatially varying
chemical potential has not been studied. In
Fig.~\ref{fig:influence-grid-points}, we also plot the effect of the number of
radial grid points on the resistivity in order to have an estimate on the
convergence of numerical errors for these quantities. We notice the effect is
stronger on the longitudinal resistivity.

\begin{figure}
  \centering
  \includegraphics[width=\textwidth]{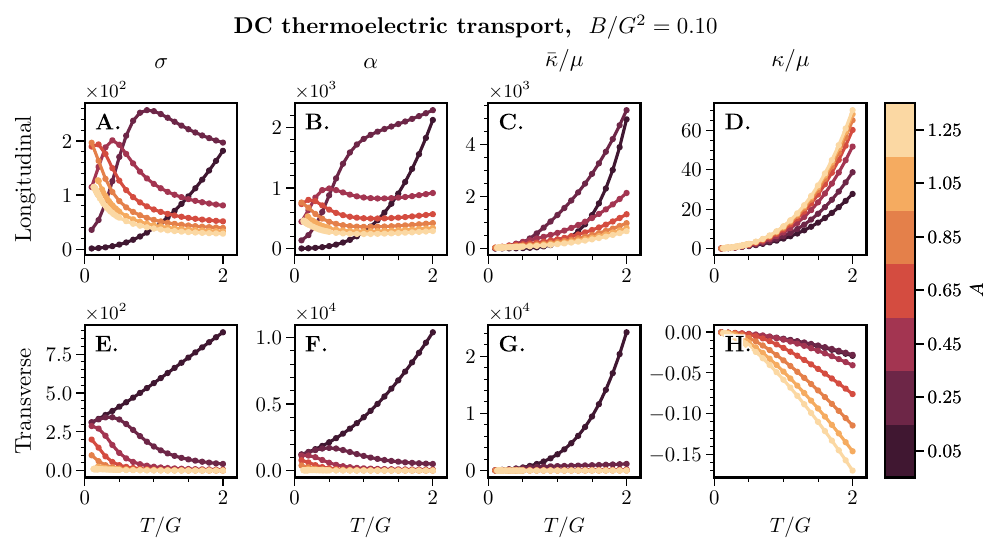}
  \caption{\footnotesize{The longitudinal and transverse Hall DC thermoelectric
      conductivities, $\sigma_{ij}, \alpha_{ij}, \bar \kappa_{ij},
      \kappa_{ij}=\bar{\kappa}_{ij}-\frac{1}{T}(\alpha\sigma^{-1}.\alpha)_{ij}$,
      computed numerically for a 2D holographic \acl{rn} strange
      metal models
      in a periodic potential $\mu_{\text{ext}}=\bmu + \frac{\bmu
      A}{2}\bigl(\cos(G
      x) + \cos(G y)\bigr)$ with $G/\bmu=\num{0.1}$ and a background
      perpendicular
      magnetic field $B/G^2=\num{0.1}$. Visually the longitudinal regime of the
      conductivities is not very enlightening as at the temperatures presented
      the
      low $A$ results are in the apparent magnetic insulator regime $\sigma_{xx}
      \sim \frac{1/\tau_0}{\frac{1}{\tau_0^2}+\omega_c^2}$ when $\omega_c \gg
  1/\tau_0$.}}
  \label{fig:thermoelectric-conductivities}
\end{figure}

The final result for the full DC thermoelectric response is presented in
Fig.~\ref{fig:thermoelectric-conductivities}.

\section{Weak lattice hydrodynamics with a magnetic field}
\label{appendix:magneto-hydro}

\subsection{DC results}
\label{appendix:subsec:dc-magneto-hydro}

This appendix provides the details of the derivation of
the relaxation rate and resistivity of a charged fluid with weak momentum
relaxation induced by a periodic chemical potential and a background
transverse magnetic field. In order to obtain the DC transport properties, we
will
drive the system in such a way that a parametrically large spatially averaged
fluid velocity is generated, which balances with the momentum relaxation to
create a steady state flow.
This method was detailed in
\cite{lucasConductivityStrangeMetal2015,lucasHydrodynamicTransportStrongly2015}
without any magnetic field and in this section we will mostly highlight the
differences which arise from adding the magnetic field contributions.

A charged fluid in $2+1$ dimensions with energy, momentum and charge
conservation admits the following conservation equations
\begin{align}
  \label{eq:hydro-conservation-eqs}
  \partial_t s + \partial_i (j_Q^i/T)     & = 0~,
  \non
  \partial_t n + \partial_i j^i           & = 0~,
  \non
  \partial_t \pi^i + \partial_j \tau^{ij} & = n E^i + B^i_{~m} j^m~,
\end{align}
where $s$, $n$ and $\pi^i$ are the entropy, charge and momentum densities. The
equation for the entropy density can be obtained from the conservation of
energy $\partial_t \epsilon + \partial_i j_E^i = j_i \partial^i \mu + j_Q^i
\frac{\partial^i T}{T}$ where $\epsilon = sT + \mu n - P$ is the energy
density and $j_E^i \equiv j_Q^i + \mu j^i$ is the energy flux upon using the
first law $\dd P=s \dd T + n \dd \mu$.
Moreover, $E^i$ is a background electric field and $B_{ij} = B_0 \epsilon_{ij}$
is the anti-symmetric strength tensor associated with the magnetic field.

Under the assumption that all length/time-scales of interest are larger than the thermalization scale,
we
can expand the heat, charge and momentum fluxes in a gradient expansion which
at linear order gives
\begin{align}
  \label{eq:hydro-constitutive-relations}
  j_Q^i     & = s T v^i -  T \alpha_Q \left(\partial^i \mu - E^i - B^{i}_{~j}
  v^j \right)  - \bar{\kappa}_Q\partial^i T + \mathcal O(\partial^2)~,
  \non
  j^i       & = n v^i   -  \sigma_Q (\partial^i \mu - E^i - B^{i}_{~j} v^j) -
  \alpha_Q \partial^i T  + \partial_j M^{ij} + \mathcal O(\partial^2)~,
  \non
  \tau^{ij} & = P  \delta^{ij} - B^{i}_{~m}M^{jm} - \eta \left( \partial^i v^j
  + \partial^j v^i - \delta^{ij} \partial_m v^m \right) + \mathcal
  O(\partial^2)~.
\end{align}
In the previous expression, we assumed the bulk viscosity is vanishing for simplicity, and that the thermodynamic equilibrium
is rotationally and parity invariant, i.e. the transport coefficients are scalars. We can
further use Lorentz invariance (inherent to our holographic model) to relate
all the thermoelectric transport coefficients to the transport coefficient
associated with charge transport at rest $\sigma_Q$ such that $\alpha_Q = -
\frac{\mu}{T} \sigma_Q$ and $\bar \kappa_Q = \frac{\mu^2}{T} \sigma_Q$. The
non-relativistic case is discussed in Appendix \ref{sec:galilean-limit}.

The conservation Eqs. \eqref{eq:hydro-conservation-eqs} in the presence of
a background electric field $E_i = \partial_i \mu_0(\bx)$ admits an
equilibrium solution in the rest frame of the fluid $v^i(\bx) = 0$ with
$T(\bx) = T_0$, $\mu(\bx) = \mu_0(\bx)$ and the hydrostatic condition
$\partial_i P = n \partial_i \mu_0$. The equilibrium values of the charge and
entropy densities are determined by the equation of state $P = P(T,\mu)$ with
$n = \frac{\partial P}{\partial \mu}$ and $s = \frac{\partial P}{\partial T}$.

Note that at equilibrium, the magnetization tensor takes a constant value
$M_{ij} = M \epsilon_{ij}$ which, due to its anti-symmetry, does not
contribute to the conservation equations and thus does not alter the
equilibrium solution.

We now consider spatially varying but static constant-in-time (DC) linear
perturbations on top of this background driven by an additional electric field $\delta E_i$, and linearize the constitutive
relations
\begin{align}
  \label{eq:hydro-linear-constitutive-relations}
  \delta j_Q^i     & = s T  \delta v^i +\mu \sigma_Q \bigl[ \left(\partial^i
    \delta \mu - \delta E^i - B^{i}_{~j} \delta v^j \right)  -
  \frac{\mu}{T}\partial^i \delta T \bigr] + \mathcal O(\partial^2)~, \non
  \delta j^i       & = n \delta v^i   - \sigma_Q \bigl[\partial^i \delta \mu -
    \delta E^i - B^{i}_{~j} \delta v^j  - \frac{\mu}{T} \partial^i \delta T
  \bigr] + \mathcal O(\partial^2)~,                  \non
  \delta \tau^{ij} & = \delta P  \delta^{ij}  - \eta \left( \partial^i \delta
  v^j + \partial^j \delta v^i - \delta^{ij} \partial_m \delta v^m \right) +
  \mathcal O(\partial^2)~.
\end{align}
In the previous expression, we did not include any fluctuation in the
magnetization since they will not contribute to the equations of motion
\textemdash{} for the reasons we highlighted previously \textemdash{} and so
we can neglect them in this procedure.
The linearized conservation equations Eqs.~\eqref{eq:hydro-conservation-eqs}
for static fluctuations take the form
\begin{align}
  \label{eq:hydro-linear-conservation-eqs}
  \partial_i \Bigl(s T  \delta v^i \Bigl) + \partial_i \Bigl[\mu \sigma_Q
  \bigl(\partial^i \delta \mu - \delta E^i - B^{i}_{~j} \delta v^j \bigr) \Bigr]
  - \partial_i \Bigl(\sigma_Q \frac{\mu^2}{T}\partial^i \delta T \bigr) & = 0~,
  \non
  \partial_i \Bigl(n \delta v^i\Bigr)   - \partial_i \Bigr[\sigma_Q
  \bigl(\partial^i \delta \mu - \delta E^i - B^{i}_{~j} \delta v^j \bigr) \Bigr]
  + \partial_i \Bigl[ \sigma_Q\frac{\mu}{T} \partial^i \delta T  \Bigr]       &
  = 0~, \non
  s \partial_i \delta T + n \partial_i \delta \mu  - \partial_i \Bigl[  \eta
    \left( \partial^i \delta v^j + \partial^j \delta v^i - \delta^{ij}
      \partial_m
  \delta v^m \right) \Bigr]                                            & =
  \non
  &\hspace{-3.5in}
  n \delta E_i +  B_{im} \left[n \delta v^m  - \sigma_Q \left(\partial^m \delta
      \mu - \delta E^m - B^m_{~j} \delta v^j   -\frac{\mu}{T} \partial^m \delta
      T
  \right) \right]                                                      ~.
\end{align}
In the last equation, we used that the charge density fluctuation term $E_i
\delta n$ cancels the pressure fluctuation terms $\delta T \partial_i s +
\delta \mu \partial_i n$ as
\begin{align}
  \delta T \partial_i s + \delta \mu \partial_i n = \delta T \chi_{sn}
  \partial_i \mu + \delta \mu \chi_{nn} \partial_i  = \underbracket{\Bigl(\delta
  T \chi_{sn} + \delta \mu \chi_{nn} \Bigr)}_{=\delta n} \partial_i \mu~,
\end{align}
remembering that since $\mu_0 = \mu$, then $\partial_i \mu = E_i=\partial_i\mu_0$.
The system of four equations \eqref{eq:hydro-linear-conservation-eqs} is therefore
closed for the four variables $\delta T, \delta \mu, \delta v^i$.

We now make the additional assumption that the translation symmetry breaking is
weak
\begin{align}
  \mu_0(\bx) = \bar{\mu}+\eps \hat{\mu}(\bx)
\end{align}
with $\eps$ a small parameter. In many steps below we shall even choose the
more specific form
\begin{align}
  \label{eq:hydro-mu-form}
  \mu_0(\bx) = \bmu + \ve \frac{\bar \mu A}{2} \Bigl[ \cos(G x) + \cos(G y)
  \Bigr]~.
\end{align}
Following the lead of \cite{lucasHydrodynamicTransportStrongly2015}, we then
Fourier transform the fluctuations and expand them in powers of $\ve$ in the
following manner
\begin{align}
  \label{eq:hydro-expansion-fluctuations-lucas}
  \begin{gathered}
    \delta \mu = \sum_{n \geq -1} \ve^n \dmuF{n}(\bk)e^{i\bk\cdot \bx}~,\quad
    \delta T = \sum_{n \geq -1} \ve^n \dTF{n}(\bk)e^{i\bk\cdot \bx}~,\\
    \delta v^i = \sum_{n \geq -2} \ve^n \dvB{n}^i + \sum_{n \geq -1} \ve^n
    \dvF{n}^i (\bk)e^{i\bk\cdot \bx}~,\\
    X(\bx) = \sum_{n \geq 0} \ve^n \barv{X}{n} + \sum_{n \geq 1}
    \frac{\ve^n}{n!}  \frac{\partial^n X}{\partial \mu \cdots \partial \mu} \, \muh{n}(\bk) e^{i \bk \cdot
    \bx}~, \quad X \in \{\mu, n, s, \eta, \sigma_Q\}~.
  \end{gathered}
\end{align}
In this expansion, we separated the zero momentum from the finite momentum
contributions where the former are collectively denoted $\bar X = \sum_{n \geq
0} \ve^n \bar{X}_{(n)}$.
In the finite momentum contributions, we defined 
$\muh{n}(\bk)$ as
the Fourier transform of $\bigl( \mu(\bx) - \bmu \bigr)^n$.
A crucial difference in our approach compared to
\cite{lucasHydrodynamicTransportStrongly2015} is the presence of the magnetic
field $B$. We notice that in the Lorenz force contribution $n \delta E_i +
B_{ij} \delta j^i$, the first term scales as $\ve^0$ in the small parameter,
while at leading order the current is expected to scale as $1/A^2 \sim
\ve^{-2}$ for a lattice, \emph{i.e.}, $\delta j^i \sim \dvB{-2} \ve^{-2}$.
Therefore, we will choose to simultaneously scale $B_0 \sim \ve^2$ so the two
terms get similar contributions, as also pointed out in \cite{Lucas:2016omy}.
The physical reason behind this choice lies in the fact that any other scaling
choice will either make the convective part or the magnetic part dominant at
leading order; however we are interested in the interplay between the
resistive longitudinal transport and magnetic transverse transport which thus
motivates us to balance these two terms.

The computation of the linear response transport of this systems is then rather
straightforward; at each order $n \geq -1$, the finite momentum projection of
the equations \eqref{eq:hydro-linear-conservation-eqs} yields a relation
between $\dmuF{n}, \dTF{n}, \dvF{n}^i$ and $\dvB{n-1}^i$. Then, the projection
of the 
equations at $\bk = 0$ and at order $\ve^{n+1}$ allows one to
solve for the remaining variables $\dvB{n-1}^i$.\footnote{Note that out of the
  4 equations, the heat and charge conservation equations have no support at
  $\bk = 0$ since they are exact divergences. Thus when projecting to zero
  momentum, we are left with the two momentum conservation equations to solve
for the two variables and thus the system is closed.} Since each moment must
vanish in the absence of an electric field, each remaining variable
$\dvB{n-1}^i$ is proportional to the driving electric field fluctuation
$\delta E_i$. We can thus relate the fluid velocity to the electric field
fluctuation as $\delta \bar{v}^i =  S^{ij} \delta E_j$ which can be translated
into a relaxation rate matrix $\tau^{-1}$ as
\begin{align}
  \label{eq:hydro-relaxation-rate-matrix}
  \bigl(\tau^{-1}\bigr)_{ij} \equiv \barv{\chipp}{0}^{-1} \Bigl( \barv{n}{0}
  \delta_{ik} + \barv{\sigma_Q}{0} B_0 \epsilon_{ik} \Bigr) S^{-1}_{kj}~, \quad
  \barv{\chipp}{0} \equiv T_0 \barv{s}{0} + \barv{\mu}{0} \barv{n}{0}~.
\end{align}
Note that the relaxation rate matrix inherits the series expansion form from
our fluctuation expansion \eqref{eq:hydro-expansion-fluctuations-lucas} as
$(\tau^{-1}) = \sum_{n \geq 1} \tau^{-1}_{(2n)} \ve^{2n}$ where the parity
invariance ensures that only even powers are present in this series. At
leading and subleading orders in $\ve$ and at leading order in $A$, this
matrix takes the form
\begin{align}
  \label{eq:hydro-relaxation-rate-leading-subleading}
  \tau_{xx}^{-1} & = \tau_0^{-1} + \Bigl(
    \frac{\barv{\sigma_{Q}}{0} B_0^2}{\barv{\chipp}{0}} + A^4 \tau_{xx,(4)}^{-1}
  \Bigr) + \mathcal O(\ve^6)~,
  \non
  \tau_{xy}^{-1} & = - \omega_c ~~\underbracket[0.140ex]{-
    \frac{B_0}{\barv{\chipp}{0}} \Bigl( \barv{n}{2} + \barv{Q}{2} + \mathcal
  O(A^4/B_0) \Bigr)}_{=\tau^{-1}_{xy,(4)}}+ \mathcal O(\ve^6)~, \non
  \barv{Q}{2}    & =- \frac{\bmu^2 A^2}{8 \barv{\chipp}{0}^2}
  \Biggl[ \barv{\cen}{0} \Bigl( \barv{\chipp}{0} \barv{\cnn}{0}
    -\barv{\sigma_Q}{0} \barv{\eta}{0} G^2 \Bigr) + T_0 \dta{sn}{\mu} \Bigl(
      \bar{n} + \frac{\barv{\chipp}{0}}{\barv{\sigma_Q}{0}}
      \susone{\sigma_Q}{\mu}
  \Bigr)  \Biggr]~,
\end{align}
with the two leading order terms are defined as
\begin{align}
  \label{eq:hydro-leading-order}
  \omega_c = \omegaccan = \frac{\barv{n}{0} B_0}{\barv{\chipp}{0}}~, \quad
  \tau_0^{-1}  = \frac{\bmu^2 A^2}{8 \barv{\chipp}{0}^3} \biggl[
    \frac{T_0^2}{\barv{\sigma_Q}{0}}\bigl( \dta{sn}{\mu} \bigr)^2 +
  \barv{\cen}{0}^2 \barv{\eta}{0} G^2 \biggr]~.
\end{align}
Here we used $\barv{\cen}{0} = T_0 \barv{\cns}{0} + \bmu \barv{\cnn}{0}$ and
$\dta{sn}{\mu} = \barv{s}{0} \barv{\cnn}{0} - \barv{n}{0} \barv{\cns}{0}$.

The resistivity matrix can then be related to the relaxation rate matrix by
using that $\delta E_m = \rho_{mn} \delta j^n$ as well as by using the
constitutive relation for $\delta j^n$ (and the solutions of our perturbative
equations) such that eventually, we find the following relation between the
subleading and the leading orders in the resistivity matrix
\begin{align}
  \label{eq:hydro-relaxation-rate-resistivity}
  \rho_{xx} & =  \frac{\tau_0^{-1}}{\omegapcan}
  -\frac{1}{\omegapcan}\biggl[\underbrace{\frac{\tau_0^{-1}}{\barv{n}{0}} \Bigl(
        2\barv{n}{2} + \barv{Q}{2} + \frac{\barv{\sigma_Q}{0} \tau_0^{-1}
    \barv{\chipp}{0}}{\barv{n}{0}} \Bigr) - A^4 \tau_{xx,(4)}^{-1}}_{\sim A^4}
  \biggr] + \mathcal O(\ve^6)~, \non
  \rho_{xy} & = -\frac{B_0}{\barv{n}{0}}  +
  \Biggl[\frac{\tau^{-1}_{xy,(4)}}{\omegapcan} + \frac{B_0}{\barv{n}{0}^2}
  \bigl( \barv{Q}{2} + 2\barv{n}{2} \bigl)  \Biggr] + \mathcal O(\ve^6)~.
\end{align}
From the expressions \eqref{eq:hydro-relaxation-rate-leading-subleading}, we
see 
the remarkable cancellation in
Eqs.~\eqref{eq:hydro-relaxation-rate-resistivity} such that the resistivity
simplifies to
\begin{align}
  \label{eq:hydro-relaxation-rate-resistivity-cancelled}
  \omegapcan \rho_{xx} & =  \tau_0^{-1} + A^4 \rho_{xx,(4)}
  + \mathcal O(\ve^6)~,
  \non
  \rho_{xy}            & = -\frac{B_0}{\barv{n}{0}} \Bigl[ 1 -
  \frac{\barv{n}{2}}{\barv{n}{0}}   \Bigr]  + \mathcal O(\ve^6) \simeq -
  \frac{B_0}{\bar n} + \mathcal O(\ve^6)~,
\end{align}
where in the last equality for $\rho_{xy}$, we recognize the remaining term as
the next order in the hydrostatic expansion of $\frac{1}{\bar n}$.

\begin{table}[t!]
  \begin{align*}
    \hspace{-.1in}
    \begin{array}{|>{\scriptstyle}l|}
      \hline
      \Gamma_{(4)} =\frac{\barv{\cee}{0} \barv{\cen}{0}^4 \barv{\eta}{0}^3 G^4
      \bmu^4}{64 \barv{\chipp}{0}^8}-\frac{\bmu^4 (\barv{\cnn}{0}
      \barv{\chipp}{0}-\barv{\cen}{0} \barv{n}{0})^4}{256 \barv{\chipp}{0}^5
      \barv{\eta}{0} G^2 \barv{\sigma_Q}{0}^2}
      \\+\frac{\bmu^4 (\barv{\cnn}{0} \barv{\chipp}{0}-\barv{\cen}{0}
        \barv{n}{0})^4 \left(\barv{\cnn}{0} \barv{\chipp}{0}^2+\barv{n}{0} (-2
      \barv{\cen}{0} \barv{\chipp}{0}+\barv{\cee}{0} \barv{n}{0})\right)}{64
      \barv{\chipp}{0}^8 G^2 \barv{\sigma_Q}{0}^3} \\
      +\frac{\barv{\cen}{0} \barv{\eta}{0} \bmu^4 (\barv{\cnn}{0}
      \barv{\chipp}{0}-\barv{\cen}{0} \barv{n}{0})^2}{64 \barv{\chipp}{0}^8
      \barv{\sigma_Q}{0}^2} \Biggl(3 \barv{\cen}{0} \barv{\cnn}{0}
        \barv{\chipp}{0}^2-2 \left(2 \barv{\cen}{0}^2+\barv{\cee}{0}
        \barv{\cnn}{0}\right) \barv{\chipp}{0} \barv{n}{0}+3 \barv{\cee}{0}
      \barv{\cen}{0} \barv{n}{0}^2\Biggr)\\
      +\frac{\barv{\cen}{0}^2 \barv{\eta}{0}^2 G^2 \bmu^4 (\barv{\cnn}{0}
      \barv{\chipp}{0}-\barv{\cen}{0} \barv{n}{0})}{64 \barv{\chipp}{0}^8
      \barv{\sigma_Q}{0}} \Biggl(2 \barv{\cen}{0}^2
        \barv{\chipp}{0}+\barv{\cee}{0}
        \barv{\cnn}{0} \barv{\chipp}{0}-3 \barv{\cee}{0} \barv{\cen}{0}
      \barv{n}{0}\Biggr)-\frac{3 \barv{\cen}{0}^2 \barv{\eta}{0}^2 G^4 \bmu^4
      \barv{\sigma_Q}{0}}{128 \barv{\chipp}{0}^5} \\
      +\frac{\barv{\eta}{0} G^2 \bmu^4}{256 \barv{\chipp}{0}^5} \Biggl(23
        \barv{\cen}{0}^4+48 \barv{\cen}{0}^3 \barv{n}{0}-4 \barv{\cen}{0}^2
        \left(12
          \barv{\cnn}{0} \barv{\chipp}{0}+9 \barv{\chipp}{0} \sustwom{n}{\mu}
          \bmu-2
        \barv{n}{0}^2+9 \barv{\chipp}{0} \sustwom{s}{\mu} T_0\right) \\
        +2 \barv{\chipp}{0}^2 \left(6 \barv{\cnn}{0}^2+8 \barv{\cnn}{0}
          (\sustwom{n}{\mu} \bmu+\sustwom{s}{\mu} T_0)+3 (\sustwom{n}{\mu}
        \bmu+\sustwom{s}{\mu} T_0)^2\right) \\
        +2 \barv{\cen}{0} \barv{\chipp}{0} (-10 \barv{\cnn}{0} \barv{n}{0}-11
          \barv{n}{0} (\sustwom{n}{\mu} \bmu+\sustwom{s}{\mu} T_0)+3
          \barv{\chipp}{0}
          (\sustwom{n}{\mu}+\susthreem{n}{\mu} \bmu+\susthreem{s}{\mu}
      T_0))\Biggr)\\
      +\frac{\bmu^4}{512 \barv{\chipp}{0}^5 \barv{\sigma_Q}{0}} \Biggl(28
        \barv{\cen}{0}^4 \barv{n}{0}^2+8 \barv{\cen}{0}^3 \barv{n}{0} \left(-7
        \barv{\cnn}{0} \barv{\chipp}{0}+8 \barv{n}{0}^2\right)
        \\ +\barv{\chipp}{0}^2 \biggl(-32 \barv{\cnn}{0}^3 \barv{\chipp}{0}+5
          (-\barv{\chipp}{0} \sustwom{n}{\mu}+\sustwom{n}{\mu} \bmu
          \barv{n}{0}+\sustwom{s}{\mu} \barv{n}{0} T_0)^2
          +12 \barv{\cnn}{0} ((\barv{\chipp}{0} \susthreem{n}{\mu}-2
            \sustwom{n}{\mu} \barv{n}{0}) (\barv{\chipp}{0}-\bmu \barv{n}{0})
            \\+\barv{n}{0} (-\barv{\chipp}{0} \susthreem{s}{\mu}+2
          \sustwom{s}{\mu} \barv{n}{0}) T_0)+4 \barv{\cnn}{0}^2 \left(3
            \barv{n}{0}^2-5
            \barv{\chipp}{0} (\sustwom{n}{\mu} \bmu+\sustwom{s}{\mu}
        T_0)\right)\biggr)\\
        +4 \barv{\cen}{0}^2 \left(7 \barv{\cnn}{0}^2 \barv{\chipp}{0}^2-40
          \barv{\cnn}{0} \barv{\chipp}{0} \barv{n}{0}^2+\barv{n}{0} \left(8
            \barv{\chipp}{0}^2 \sustwom{n}{\mu}+3 \barv{n}{0}^3-13
            \barv{\chipp}{0}
        \barv{n}{0} (\sustwom{n}{\mu} \bmu+\sustwom{s}{\mu} T_0)\right)\right)
        \\
        +4 \barv{\cen}{0} \barv{\chipp}{0} \Bigl(32 \barv{\cnn}{0}^2
          \barv{\chipp}{0} \barv{n}{0}+3 \barv{n}{0} (-((\barv{\chipp}{0}
              \susthreem{n}{\mu}-2 \sustwom{n}{\mu} \barv{n}{0})
              (\barv{\chipp}{0}-\bmu
            \barv{n}{0}))  \\
            +\barv{n}{0} (\barv{\chipp}{0} \susthreem{s}{\mu}-2
              \sustwom{s}{\mu}
          \barv{n}{0}) T_0)+2 \barv{\cnn}{0} \left(-4 \barv{\chipp}{0}^2
            \sustwom{n}{\mu}-3 \barv{n}{0}^3+9 \barv{\chipp}{0} \barv{n}{0}
      (\sustwom{n}{\mu} \bmu+\sustwom{s}{\mu} T_0)\right)\Bigr)\Biggr)~, \\[2em]
      \hline
    \end{array}
  \end{align*}
  \caption{\footnotesize{
      The %
      first subleading corrections of the momentum relaxation rate
      $\Gamma_{(4)}$
      obtained from hydrodynamics in the presence of a magnetic field and a weak perturbative
      lattice. The result is specific for a cosine lattice of the form sourced
      by
      $\mu(\bx) =\bar{\mu}(1+\frac{A}{2}(\cos(Gx)+\cos(Gy)))$ and a conformal
      liquid
  where the bulk viscosity is assumed to be zero.}}
  \label{tab:summary-Gamma4}
\end{table}

\subsection{AC results}

One question which arises from the previous subsection is whether the
relaxation rate matrix $\tau^{-1}$ defined through
Eq.~\eqref{eq:hydro-relaxation-rate-matrix} (and equally more generally
defined by Eq.~\eqref{eq:tau-def}) actually accounts for the position of poles
in the AC spectrum of the conductivity (or any other response function which
overlaps with momentum). To answer this, we will reconsider our conservation
equations \eqref{eq:hydro-conservation-eqs} at finite frequency $\omega =
\sum_{n \geq 1} \omega_{(2n)} \ve^{2n}$, using the local thermodynamic
equilibrium to relate the conserved quantities fluctuations with their
internal variables to lowest order in gradients
  \begin{align}
    \delta n     & = \cnn \delta \mu + \cns \delta T~, \non
    \delta s     & = \cns \delta \mu + \css \delta T~, \non
    \delta \pi^i & = (\bar s T_0 + \overline{\mu n}) \delta v^i~,
  \end{align}
where the various susceptibilities so-introduced can further be expanded
following \eqref{eq:hydro-expansion-fluctuations-lucas}. In this section, we
will also introduce an external thermal drive $\delta \zeta_i$ by shifting
$\partial^i T \to \partial^i T - \delta \zeta_i$ in the constitutive relations
\eqref{eq:hydro-constitutive-relations} and by introducing a new term $\bar{n}
\delta E^i \to \bar{n} \delta E^i + \bar s \delta \zeta^i$ in the right-hand
side of the linearized momentum equation in
\eqref{eq:hydro-linear-conservation-eqs}. This term will prove useful later on
to extract the thermoelectric conductivities in the hydrodynamic regime.

We can once again solve order by order the finite momentum equations in order
to deduce a zero-momentum equation for $\delta v^i$ of the form
$S_{ij}(\omega) \delta \bar{v}^j = R^E_{ij} \delta E^j + R^Z_{ij} \delta
\zeta^j$ which, expanded, becomes
\begin{align}
  \Bigl(S^{(0)}_{ij}(\omega) + \ve^2 S^{(2)}_{ij}(\omega) & + \ldots \Bigr)
  \Bigl( \dvB{-2}^j \ve^{-2} + \dvB{0}^j + \ldots \Bigr) =  \\
  & \Bigl(R^{E, (0)}_{ij} + R^{E, (2)}_{ij} \ve^2 + \ldots \Bigr) \delta E^j +
  \Bigl(R^{Z, (0)}_{ij} + R^{Z, (2)}_{ij} \ve^2 + \ldots \Bigr) \delta \zeta^j~.
  \nonumber
\end{align}
The expansion in $\omega$ can then be obtained by solving $\det P(\omega) = 0$
order by order. At leading order, we find the canonical position of the Drude
cyclotron mode at the canonical cyclotron frequency with the Drude relaxation rate
\begin{align}
  \label{eq:leading-order-pole-hydro-ac}
  \omega_{(2)} = \underbracket[0.140ex]{\omegaccan}_{\sim B} - i
  \underbracket[0.140ex]{\tau_0^{-1}}_{\sim A^2}~.
\end{align}
The sub-leading correction $\omega_{(4)}$ is  more involved but can
still be computed and we obtain a real (cyclotron) and imaginary (relaxation)
corrections
\begin{align}
  \label{eq:corrections-pole-hydro-ac}
    \omega_{(4)}   & = \omega_{(4,2)} A^2 B - i
    \underbracket[0.140ex]{\Bigl(\frac{\barv{\sigma_{Q}}{0}}{\barv{\chipp}{0}} B^2 + \omega_{(4,4)} A^4
    \Bigr)}_{{= \gamma + \Gamma_{(4)}}}~ 
\end{align}
where in the imaginary part we recognize the relaxation scale $\gamma$. The remaining term $\omega_{(4,4)}A^4 \equiv \Gamma_{
(4)}$ introduced in Eq.\eqref{eq:final-cyclotron-drude} by definition. Its full expression is given in Table \ref{tab:summary-Gamma4}. 

The real part -- the cyclotron frequency shift --- is given by
\begin{align}    
    \omega_{(4,2)} & = \frac{\bar \mu^2 \barv{n}{0} (\barv{\cen}{0})^2}{8 \,
    \barv{\chipp}{0}^6} \Bigl[ T_0^2 \barv{\css}{0} + \bar \mu^2 \barv{\cnn}{0} + 2
    \bar \mu T_0 \barv{\cns}{0} \Bigr] \barv{\eta}{0}^2 G^2  \\
    & + \frac{1}{\barv{\sigma_{Q}}{0}^2 G^2} \frac{\bar \mu^2 T_0^4
    \barv{n}{0}}{8 \, \barv{\chipp}{0}^6} \Bigl[ \barv{s}{0}^2 \barv{\css}{0} +
    \barv{n}{0}^2 \barv{\cnn}{0} + 2 \barv{n}{0} \barv{s}{0} \barv{\cns}{0} \Bigr]
    + \frac{\bar \mu^2 \barv{\cen}{0}}{4 \barv{\chipp}{0}^3}
    \barv{\sigma_{Q}}{0} \barv{\eta}{0} G^2 \non
    & - \frac{\barv{\eta}{0}}{\barv{\sigma_{Q}}{0}} \frac{T_0^2 \barv{n}{0}
    \bar \mu^2 (\barv{\cen}{0}) (\dta{sn}{\mu})}{4 \, \barv{\chipp}{0}^6} \Bigl[
    \barv{n}{0} \susone{\eps}{T} - \barv{s}{0} \susone{\eps}{\mu} \Bigr] - \frac{T_0
    \bmu^2 \dta{sn}{\mu}}{4 \,\barv{\chipp}{0}^2 \barv{\sigma_Q}{0}}
    \susone{\sigma}{\mu} \non
    & +  \frac{\bar \mu^2}{8 \, \barv{\chipp}{0}^3} \Bigl[ T_0 \barv{\chipp}{0}
      (\barv{n}{0} \sustwom{s}{\mu} - \barv{s}{0} \sustwom{n}{\mu}) +
      \barv{\cen}{0} ( T_0 \dta{sn}{\mu} + \barv{\chipp}{0} \barv{\cnn}{0} ) + 4
    \barv{n}{0} T_0 \dta{sn}{\mu}\Bigr]~.\nonumber
\end{align}
Importantly, we can easily notice that $\omega_{(4,2)}$ takes on a very
different form from the ``relaxation rate'' matrix element
$\tau^{-1}_{xy,(4)}$ in Eq.~\eqref{eq:hydro-relaxation-rate-leading-subleading}. This shows that for any finite frequency response in hydrodynamics with weakly broken (translational) symmetry, one must improve upon the formalism of \cite{lucasConductivityStrangeMetal2015,lucasHydrodynamicTransportStrongly2015} as in Appendix \ref{appendix:subsec:dc-magneto-hydro} by including a perturbative finite frequency response as explained here. This point has also been made in the context of charge density waves in \cite{armasApproximateSymmetriesPseudoGoldstones2022,amorettiHydrodynamicMagnetotransportCharge2021,amorettiHydrodynamicMagnetotransportHolographic2021}.

\subsection{Interpretation in terms of non-dissipative corrections}
\label{sec:connection-blaise-shukla}

In this subsection, we will re-interpret the various observations of the first
two subsections in terms of the observations in \cite{armasApproximateSymmetriesPseudoGoldstones2022} which followed from studies on charge density waves in \cite{armasHydrodynamicsChargeDensity2020,amorettiHydrodynamicMagnetotransportCharge2021,amorettiHydrodynamicMagnetotransportHolographic2021,armasHydrodynamicsPlasticDeformations2023a} and was the foundation for \cite{gouterauxDrudeTransportHydrodynamic2023,Gouteraux:2024adm}. Summarizing
briefly in the language of \cite{gouterauxDrudeTransportHydrodynamic2023}, these works show
that in models with disorder
generated by a scalar operator, one should carefully take into account
\emph{non-hydrostatic} and \emph{non-dissipative} corrections to the
constitutive relations for the currents when one computes finite frequency fluctuations. \footnote{In \cite{gouterauxDrudeTransportHydrodynamic2023}, the authors focused on the
  application to Galilean invariant systems for which $\lambda_n = 0$. However,
  we will be more interested in the applications of Lorentz invariant systems
  (such cases arise generically in holographic models) for which instead $T
\lambda_n + \mu \lambda_s = 0 = \lambda_\epsilon$.}
  \label{eq:blaise-shukla-corrections}
  \begin{align}
    \delta j^i   & \to \delta j^i + \ve^2 \lambda_n \delta v^i~,               \non
    \delta j_Q^i & \to \delta j^i + \ve^2  T_0 \lambda_s \delta v^i~,          \non
    \delta \pi^i & \to \chi_{\pi\pi} \delta v^i + \ve^2 \lambda_\pi \delta v^i~.
  \end{align}
The distinction between various corrections is made at the level of the entropy
divergence equation: {hydrostatic} corrections do not generate
dissipation and thus do not source entropy \textemdash{} they are made of
derivative corrections to the thermodynamics. In our model, these are
exemplified by the effective zero-momentum charge density $\int \dd x \dd y
\bar n(x) = \bar{n}_{(0)} + \ve^2 \bar{n}_{(2)} + \ldots$ where $\bar{n}_{(2)}
= \frac{1}{8} \chi_{nn} A^2 \bmu^2$ for the cosine chemical potential modulation Eq.~\eqref{eq:hydro-mu-form}. On the other hand, dissipative
corrections such as $\sigma_Q \partial^i \mu$ or $\eta \partial^i v^j$
contribute to the entropy divergence equations. However, the new corrections $\lambda_i$ introduced
in Eq.~\eqref{eq:blaise-shukla-corrections} belong to neither categories: 
they are formally present in the equation but their contribution to entropy-production cancels
exactly.

The translational symmetry breaking introduced by a scalar operator in \cite{armasApproximateSymmetriesPseudoGoldstones2022,armasHydrodynamicsChargeDensity2020,amorettiHydrodynamicMagnetotransportCharge2021,amorettiHydrodynamicMagnetotransportHolographic2021,armasApproximateSymmetriesPseudoGoldstones2022,armasHydrodynamicsPlasticDeformations2023a} and \cite{gouterauxDrudeTransportHydrodynamic2023} also shows up as a source in the fluctuations of the momentum conservation equation
\begin{align}
  \label{eq:momentum-equation-hydro}
  \partial_t \delta \pi^i + \partial_j \delta \tau^{ij} = - \Gamma \delta \pi^i
  + \bar n E^i + F^{ik} \delta j_k - \ve^2 \lambda_n \left( \partial^i \mu - E^i
  - F^{ik} \delta v_k \right) - \ve^2 \lambda_s \partial^i T~.
\end{align}
where the key part is that the coefficients of the source terms corresponding to gradients in the thermodynamic potentials are the same $\lambda_i$ as introduced in Eq.~\eqref{eq:blaise-shukla-corrections}. This follows from the single thermodynamic free energy coupled to background sources for the momentum current (the spatial metric) and for the conserved current associated with the scalar operators.
In the previous expression, $\Gamma$ is the effective relaxation rate due to
the scalar operator contribution in the model of
\cite{gouterauxDrudeTransportHydrodynamic2023} and would, in the periodic chemical potential model here,
match to $\Gamma =\ve^2 \tau_0^{-1} + \omega_{(4,4)} \ve^4 A^4 + \ldots$. In
the presence of a magnetic field, the first term $F^{ik} \delta j_k$ yields a
contribution $\ve^2 B \epsilon^{ik} (\bar{n}_{(0)} + \ve^2 \bar{n}_{(2)} +
\ve^2 \lambda_n)\delta v_k$ while an extra $\ve^2 B \lambda_n \epsilon^{ik}
\delta v_k$ comes from the correction to $\partial^i \mu$. This argument can
be made more precise by combining the constitutive relations for the currents
\eqref{eq:blaise-shukla-corrections} with the momentum conservation equation
\eqref{eq:momentum-equation-hydro} projected at zero momentum\footnote{To
  simplify the expressions, we only focus on the charge sector here and ignore
the external thermal sources required to compute $\alpha$, $\bar \kappa$.}
\begin{align}
  (- i \omega + \Gamma) \delta \pi^i = (\bar n + \ve^2 \lambda_n) E^i + B
  \epsilon^{ik} \Bigl[(\bar n + 2 \ve^2 \lambda_n) \delta v_k + \sigma_Q E_k + B
  \epsilon_{kl} \delta v^l\Bigr]~,
\end{align}
which can be rewritten in matrix form in a similar way as \eqref{eq:tau-def}
\begin{align}
  \begin{bmatrix}
    \Gamma + \gamma - i \omega & -\omega_c                  \\
    \omega_c                   & \Gamma + \gamma - i \omega
  \end{bmatrix} \cdot \delta v =
  \begin{bmatrix}
    \frac{\chi_{\pi j}}{\chipp}           & \frac{\sigma_Q B}{\chipp} \\
    -\frac{\sigma_Q B}{\chipp} & \frac{\chi_{\pi j}}{\chipp}
  \end{bmatrix} \cdot E~,
\end{align}
with
\begin{align}
  & \omega_c = \frac{\bar{n}_{(0)} + \ve^2 \bar{n}_{(2)} + 2 \ve^2
  \lambda_n}{\bar{\chi}_{\pi \pi, (0)} + \ve^2 \bar{\chi}_{\pi \pi, (2)} + \ve^2
  \lambda_\pi} \ve^2 B~, &  & \quad \chipp = \barv{\chipp}{0} + \ve^2
  \barv{\chipp}{2} + \ve^2 \lambda_\pi                                     \non
  &  \quad \gamma = \ve^4 \frac{\sigma_Q
  B^2}{\chipp}~, && \quad \chi_{\pi j} = \barv{n}{0} + \ve^2 \barv{n}{2} + \ve^2 \lambda_n
\end{align}
{Note that while usually $\chi_{\pi j}$  is simply the thermodynamic charge density at lowest order, here we also must take into account the non-hydrostatic non-dissipative corrections, introduced in Eq.\eqref{eq:blaise-shukla-corrections}.
From this simple Drude calculation, we can see the effect of the
non-dissipative corrections on the cyclotron frequency and the Drude weight.
Combining this result with the constitutive relations, 
one extracts the Drude conductivities
\begin{align}
  \sigma_{ij}(\omega) = \sigma_Q \delta_{ij} + \frac{(\chipp)^{-1}}{\bigl(
  \Gamma + \gamma - i \omega \bigr)^2 + \omega_c^2}
  \begin{bmatrix}
    \chi_{\pi j}           & \sigma_Q B \\
    -\sigma_Q B & \chi_{\pi j}
  \end{bmatrix} \cdot
  \begin{bmatrix}
    \Gamma + \gamma - i \omega & \omega_c                   \\
    -\omega_c                  & \Gamma + \gamma - i \omega
  \end{bmatrix} \cdot
  \begin{bmatrix}
    \chi_{\pi j}           & \sigma_Q B \\
    -\sigma_Q B & \chi_{\pi j}
  \end{bmatrix}
\end{align}
Note that the first two matrices actually commute so this can be rewritten as
\begin{align}
  \sigma_{ij}(\omega) = \sigma_Q \delta_{ij} +\frac{1}{\bigl( \Gamma + \gamma -
  i \omega \bigr)^2 + \omega_c^2}
  \begin{bmatrix}
    \Gamma + \gamma - i \omega & \omega_c                   \\
    -\omega_c                  & \Gamma + \gamma - i \omega
  \end{bmatrix} \cdot
  \begin{bmatrix}
    \omega_p^2 - \sigma_Q \gamma   & 2 \sigma_Q \frac{B \chi_{\pi j}}{\chipp} \\
    -2 \sigma_Q \frac{B \chi_{\pi j}}{\chipp} & \omega_p^2 - \sigma_Q \gamma
  \end{bmatrix}
\end{align}
{where the corrected Drude weight is $\omega_p^2 = \frac{\chi_{\pi j}^2}{\chipp}$.}
More generally, by writing the conductivities as complex combinations $Z_c=Z_{xx}+iZ_{xy}$, one has the convenient form (see
\cite{amorettiMagnetothermalTransportImplies2020,gouterauxDrudeTransportHydrodynamic2023})
\begin{align}
  \label{eq:gouteraux-shukla-conductivities}
    \sigma_c(\omega) & = \sigma_Q + i\frac{\bigl( \chi_{\pi j} + i B \ve^2 \sigma_Q
    \bigr)^2}{\chipp} \frac{1}{\omega + \omega_c + i (\Gamma + \gamma)}~,
    \non
    \alpha_c(\omega) & = \alpha_Q + i\frac{\bigl( \chi_{\pi j} + i B \ve^2 \sigma_Q
    \bigr)\bigl( \chi_{\pi j_Q} + i B \ve^2 \alpha_Q \bigr)}{\chipp} \frac{1}{\omega +
      \omega_c
    + i (\Gamma + \gamma)}~, \non
    \sigma_c(\omega) & = \bar \kappa_Q + i T \frac{\bigl(  \chi_{\pi j_Q} + i B \ve^2
    \alpha_Q \bigr)^2}{\chipp} \frac{1}{\omega + \omega_c + i (\Gamma +
    \gamma)}~,
  \end{align}
which isolates only one cyclotron pole. In this form the quantities $n,s,\chipp$
which contain the hydrostatic plus non-dissipative correction $ \chi_{\pi j_Q} = \barv{s}{0} +
\ve^2 \barv{s}{2} + \ve^2 \lambda_s$, etc... are readily extracted.

\bigskip

At a first glance, it would
seem that the new corrections introduced only renormalize the values of the
susceptibilities $\chi_{\pi\pi}$, $\chi_{\pi j}$ and $\chi_{\pi j_{Q}}$ which
define the overlap between the momentum current and the charge, heat and
momentum currents, in much the same way that the hydrostatic corrections do.
One would simply expect these to
be the ``observed'' susceptibilities by any lab experiment and would simply
form the more physical quantities as opposed to the bare homogeneous
quantities. However, we see that the two non-hydrostatic contributions in the momentum
equation which lead to an extra factor of $2$ in the cyclotron frequency spoil
this line of reasoning for the non-hydrostatic corrections. In essence, a
careful measurement of both the cyclotron frequency and the Drude weight (from
fitting the AC conductivity) would be able to discriminate between $\lambda_n$
and $n_{(2)}$.
An additional consequence of this factor of $2$, highlighted already in
\cite{gouterauxDrudeTransportHydrodynamic2023}, is how it immediately leads to
the exact cancellation we also observed in the Hall coefficient $R_H =
\rho_{xy}/B$. In a Drude regime, this expression reduces to $R_H =
-\frac{\omega_c}{\omega_p^2} \ve^{-2} B^{-1}$ where one should now use the
\emph{corrected} Drude weights and cyclotron frequency. This 
simplifies at subleading order as
\begin{align}
    R_H & = -\frac{\bar{n}_{(0)} + \ve^2 \bar{n}_{(2)} + 2 \ve^2
    \lambda_n}{(\bar{n}_{(0)} + \ve^2 \bar{n}_{(2)} + \ve^2 \lambda_n)^2}~,
    \non
    & \simeq -\Bigl[ \frac{1}{\bar{n}_{(0)}} + \ve^2
      \frac{\bar{n}_{(2)}}{\bar{n}_{(0)}^2} + 2 \ve^2
    \frac{\lambda_n}{\bar{n}_{(0)}^2} \Bigr] \Bigl[ 1 - 2 \ve^2
      \frac{\bar{n}_{(2)}}{\bar{n}_{(0)}^2} - 2 \ve^2
    \frac{\lambda_n}{\bar{n}_{(0)}^2} + \ldots \Bigr]~, \non
    & = -\frac{1}{\bar{n}_{(0)}} +\ve^2 \frac{\bar{n}_{(2)}}{\bar{n}_{(0)}^2} +
    \ldots~.
  \end{align}
In the last equation, we recognize the result we also obtained in
Eq.~\eqref{eq:hydro-relaxation-rate-resistivity-cancelled} where the usual
relation between charge density and Hall coefficient is only corrected through
the hydrostatic contributions.

We emphasize, however, that so far we have only summarized and re-contextualized the results of
\cite{armasApproximateSymmetriesPseudoGoldstones2022,gouterauxDrudeTransportHydrodynamic2023} within our own framework, but
we have yet to 
show that it applies to our model. The building blocks of the two models
are quite different; in \cite{armasApproximateSymmetriesPseudoGoldstones2022,gouterauxDrudeTransportHydrodynamic2023} and the related works \cite{armasHydrodynamicsChargeDensity2020,amorettiHydrodynamicMagnetotransportCharge2021,amorettiHydrodynamicMagnetotransportHolographic2021,armasApproximateSymmetriesPseudoGoldstones2022,armasHydrodynamicsPlasticDeformations2023a} the extra coefficients $\lambda_i$ appear naturally as additional coefficients in the constitutive relations sourced by an extra scalar current. That current is not present here.  
The general framework in \cite{Gouteraux:2024adm} implies that it should, but a concrete deductive argument to account for the specific $\lambda_i$ dependence of the sources in Eq.~\eqref{eq:momentum-equation-hydro} is not known at present.
Using instead the effective Drude model \eqref{eq:gouteraux-shukla-conductivities} as a defining expression and comparing to the naive susceptibilities including the hydrostatic corrections, we can
immediately extract the values of $\lambda_n$ and $\lambda_\pi$, keeping in
mind that $\lambda_s$ is constrained by Lorentz invariance in our model. To do
so, we use the hydrodynamic setup of the previous subsection to extract the
thermoelectric conductivities $\sigma_{ij} = \frac{\delta j_i}{\delta E^j}$,
$\alpha_{ij} = \frac{\delta j_i}{\delta \zeta^j}$ and $\bar \kappa_{ij} =
\frac{\delta j_{Q,i}}{\delta \zeta^j}$. We can then take the DC limit $\omega
\to 0$ of these conductivities and project onto the sub-leading order $\ve^0$
(where the non-hydrostatic corrections first contribute), such that we can
compare the so-obtained expressions with their formal equivalent
\eqref{eq:gouteraux-shukla-conductivities}. Since one of the coefficients is
constrained, we can solve for $\lambda_n$ and $\lambda_\pi$ using $\sigma$ and
$\alpha$, and note as a consistency check that the match for $\bar \kappa$
should be, and is, automatically verified. The expressions for the
coefficients are then
\begin{align}
  \label{eq:expressions-nhs-coeffs}
    \lambda_n & = \frac{\bmu^2 A^2}{8 \barv{\chipp}{0}^2} \Bigl[
      \barv{\sigma_Q}{0} \barv{\eta}{0} G^2 \barv{\cen}{0} - (\barv{\cen}{0} + 2
      \barv{n}{0})(\barv{\chipp}{0} \barv{\cnn}{0} - \barv{n}{0}
    \barv{\cen}{0})\Bigr]~,                                          \non
    \lambda_s & = -\frac{\bmu^3}{8 T_0 \barv{\chipp}{0}^2} A^2 \Bigl[
      \barv{\sigma_Q}{0} \barv{\eta}{0} G^2 \barv{\cen}{0} - (\barv{\cen}{0} + 2
      \barv{n}{0})(\barv{\chipp}{0} \barv{\cnn}{0} - \barv{n}{0}
    \barv{\cen}{0})\Bigr]~,                                     \non
    \lambda_\pi & = \frac{\bmu^2 A^2}{8 \barv{\chipp}{0}^4} \Bigl[
      \barv{\chipp}{0}^3 \barv{\cen}{0}^2 - 2 \barv{\chipp}{0}^4 \barv{\cnn}{0}
      -
      \barv{\cen}{0}^2 \barv{\cee}{0}  \barv{\eta}{0} G^2
      \\
      & - 2\frac{\barv{\cen}{0}  \barv{\eta}{0} G^2}{\barv{\sigma_Q}{0}}
      (\barv{\chipp}{0} \barv{\cnn}{0} - \barv{n}{0}
      \barv{\cen}{0})(\barv{\chipp}{0} \barv{\cen}{0} - \barv{n}{0}
      \barv{\cee}{0})
      \non
      & - \frac{\bigl( \barv{\chipp}{0} \barv{\cnn}{0} - \barv{n}{0}
        \barv{\cen}{0} \bigr)^2 \bigl( \barv{\chipp}{0}^2 \barv{\cnn}{0} - 2
          \barv{n}{0} \barv{\chipp}{0} \barv{\cen}{0} + \barv{n}{0}^2
          \barv{\cee}{0}
    \bigr)}{\barv{\sigma_Q}{0}^2 G^2} \Bigr]~.\nonumber
  \end{align}
In the previous expression, we introduced for simplicity the more 
standard notations $\barv{\cee}{0} \equiv T \frac{\partial \epsilon}{\partial
T} + \mu \frac{\partial \epsilon}{\partial T}$ and $\barv{\cen}{0} \equiv
\frac{\partial \epsilon}{\partial \mu}$.

From these expressions, we can compute the corrected cyclotron frequency
\eqref{eq:final-cyclotron-drude} and we found perfect agreement at orders
$\ve^2$ and $\ve^4$ with the expression we previously derived in
Eqs.~\eqref{eq:leading-order-pole-hydro-ac} and
\eqref{eq:corrections-pole-hydro-ac}.

\subsection{Galilean limit}
\label{sec:galilean-limit}

These results can also be derived in the Galilean non-relativistic limit, where the current flow directly corresponds to the momentum flow (up to the unit of charge
$j^i = \pi^i$. In terms of the more general hydrodynamic formalism, this means that $\sigma_Q \to 0$ and $\alpha_Q \to 0$ while $\bar
\kappa_Q$ remains finite. To connect to a
setting more familiar to condensed matter experiments. Under such assumption,
the heat current diffuses through a gradient $j_Q^i \sim s T v^i - \bar
\kappa_Q \partial^i T$ while the charge current keeps its Drude-like form $j^i
\sim n v^i$. Following through the same steps as in the previous sections, we
can derive the Drude pole and its corrections defined in
Eq.~\eqref{eq:final-cyclotron-drude}
  \begin{align}
    \Gamma   & = \eps^2 \tau_0^{-1} + \eps^4 A^4 \tau_4^{-1}~, \non
    \omega_c & = \frac{\barv{n}{0} + \eps^2 \barv{n}{2} + 2 \eps^2
    \lambda_n}{\barv{\chipp}{0} + \eps^2 \barv{\chipp}{2}} \eps^2 B~.
  \end{align}
Let us first start by noticing that as expected, in the Galilean limit, there
is no extra magnetic contribution $\gamma =
\frac{\barv{\sigma_Q}{0}}{\barv{\chipp}{0}} B^2 \eps^4$. The rest of the
results retain however the same form as in the Lorentz-invariant case, with
the added caveat that the constraint on non-hydrostatic coefficients is no
longer $T_0 \lambda_s + \bmu \lambda_n = 0$ but instead only $\lambda_n$. The
detailed contribution of the constituents to these expressions can be computed
straightforwardly and yield
\begin{align}
  \tau_0^{-1} = \frac{\bmu^2 A^2}{8 \barv{n}{0}^2 \barv{\chipp}{0}} \Biggl[
    \barv{\cnn}{0}^2 \barv{\eta}{0} G^2 + \frac{\bigl( \barv{n}{0}
        \barv{\cen}{0}
  - \barv{\cnn}{0} \barv{\chipp}{0} \bigr)^2}{T_0 \barv{\kappa_Q}{0}} \Biggr]
\end{align}
for the leading order momentum relaxation rate and
  \begin{align}
    \lambda_s & = - \frac{\bmu^2 A^2}{8 \barv{n}{0}^2 \barv{\kappa_Q}{0} T_0}
    (\barv{n}{0} \barv{\cen}{0} - \barv{\cnn}{0} \barv{\chipp}{0}) \bigl(
      \barv{\cnn}{0} \barv{\kappa_Q}{0} +\barv{n}{0} \frac{\partial \bar
    \kappa_Q}{\partial \mu} \bigr)\non
    \lambda_\pi & = -\frac{A^2 G^2 \barv{\eta}{0}^2 \bmu^2 \barv{\cnn}{0}^3}{8
    \barv{n}{0}^4}-\frac{A^2 \barv{\eta}{0} \bmu^2 \barv{\cnn}{0} (\barv{n}{0}
    \barv{\cen}{0}-\barv{\cnn}{0} \barv{\chipp}{0})^2}{4 \barv{\kappa_Q}{0}
    \barv{n}{0}^4 T_0}\\
    & + \frac{A^2 \bmu^2 \barv{\cnn}{0} (3 \barv{\cnn}{0} \barv{\chipp}{0}-2
    \barv{n}{0} (\barv{n}{0}+\barv{\cen}{0}))}{8 \barv{n}{0}^2}\nonumber\\
    &-\frac{A^2 \bmu^2 (\barv{n}{0} \barv{\cen}{0}-\barv{\cnn}{0}
      \barv{\chipp}{0})^2 \left(\barv{n}{0}^2 \barv{\cee}{0} -2 \barv{n}{0}
    \barv{\chipp}{0} \barv{\cen}{0}+\barv{\cnn}{0} \barv{\chipp}{0}^2\right)}{8
    G^2 \barv{\kappa_Q}{0}^2 \barv{n}{0}^4 T_0^2} \nonumber
  \end{align}
for the non-hydrostatic corrections.
\bibliographystyle{custom-Chagnet}
\bibliography{refs_magnetotransport}

\end{document}